\documentclass[%
reprint,
superscriptaddress,
amsmath,amssymb,
aps,
prb,
english,
]{revtex4-1}
\usepackage{amsmath}
\usepackage{amsfonts}
\usepackage{graphicx}
\usepackage{appendix}
\usepackage{amssymb}
\usepackage{bm}
\usepackage{color}
\usepackage{natbib}
\usepackage{nccmath}
\usepackage{array}
\usepackage{graphicx}
\usepackage{dcolumn}
\usepackage{bm}
\usepackage{float}
\usepackage[bookmarks=true,colorlinks,linkcolor=blue,urlcolor=blue,citecolor=blue]{hyperref}
\usepackage[mathlines]{lineno}%
\usepackage[english]{babel}
\usepackage{textcomp}

\newcommand{\beq}{\begin{equation}}
	\newcommand{\eneq}{\end{equation}}
\newcommand{\be}{\begin{equation}}
	\newcommand{\ee}{\end{equation}}
\newcommand{\bea}{\begin{eqnarray}}
	\newcommand{\eea}{\end{eqnarray}}

\renewcommand{\cite}[1]{[\onlinecite{#1}]}

\usepackage{multirow}
\usepackage{wasysym}


\begin{document}
	\title{Multi-particle scattering and breakdown of the Wiedemann-Franz law at a junction of $N$ interacting 
		quantum wires}
	\author{Domenico Giuliano}  \email{domenico.giuliano@fis.unical.it}
	\affiliation{Dipartimento di Fisica, Universit\`a della Calabria, Arcavacata di Rende I-87036, Cosenza, Italy}
	\affiliation{I.N.F.N. - Gruppo collegato di Cosenza, Arcavacata di Rende I-87036, Cosenza, Italy}
	\author{Andrea Nava}
	\affiliation{Dipartimento di Fisica, Universit\`a della Calabria, Arcavacata di Rende I-87036, Cosenza, Italy}
	\affiliation{I.N.F.N. - Gruppo collegato di Cosenza, Arcavacata di Rende I-87036, Cosenza, Italy}
	\author{Reinhold Egger}
	\affiliation{Institut f\"ur Theoretische Physik, Heinrich-Heine Universit\"at, D-40225 D\"usseldorf, Germany}
	\author{Pasquale Sodano}
	\affiliation{I.N.F.N., Sezione di Perugia, Via A. Pascoli, I-06123, Perugia, Italy}
	\author{Francesco Buccheri} \email{buccheri@hhu.de}
	\affiliation{Institut f\"ur Theoretische Physik, Heinrich-Heine Universit\"at, D-40225 D\"usseldorf, Germany}

	\date{\today}
	
	\begin{abstract}
		We analyze the charge and thermal transport at a junction of interacting quantum wires close to 
		equilibrium. Within the framework of Tomonaga-Luttinger liquids, we compute the thermal conductance for 
		a wide class of boundary conditions and detail the physical processes leading to the breakdown of the Wiedemann-Franz 
		law at the junction. We show how connecting external reservoirs to the quantum wires affects the conductance tensors 
		close to the various fixed points of the phase diagram of the junction. We therefore distinguish two types of violation 
		of the Wiedemann-Franz law: a ''trivial'' one, independent of the junction dynamics and arising from the breakdown of 
		the Fermi-liquid picture in the wire, and a junction-related counterpart, arising from multi-particle scattering processes at the junction.
	\end{abstract}
	\maketitle
	
	\section{Introduction} 
	\label{sec:Introduction}

	Junctions of interacting quantum wires (QWs), realized with both spinless
	\cite{Chamon2003,Oshikawa2006,Lal2002,Chen2002,Bellazzini2006}, or
	spinful systems \cite{Hou2012,Giuliano2015,Shi2016},  
	have continuously attracted  the attention of physicists, in that they can be regarded  as the simplest components of a quantum circuit.
	Furthermore, a plethora of unconventional phases can be realized at a pertinently engineered junction, which correspond to attractive fixed points in the
	boundary phase diagram of the system at which Landau's Fermi liquid paradigm breaks down.
	There are indeed various reasons for the  emergence of non Fermi liquid phases, related to the peculiar nature of the elementary excitations in effectively
	one-dimensional interacting electronic systems, to the topology of the junction, or to the dynamics of localized excitations emerging at the  junction  itself.
	In fact, the low-lying elementary excitations in an interacting fermionic system in one dimension are not particles and holes, but instead collective bosonic modes,
	whose dynamics is encoded in the  Tomonaga-Luttinger liquid  paradigm \cite{Tomonaga1955,Luttinger1963}. 
	The loss of integrity of particle and hole excitations formally corresponds to the description of tunneling processes at the junction  in terms of operators that are nonlinear functionals of the Tomonaga-Luttinger liquid fields. The corresponding scaling dimensions continuously depend on the ``bulk'' interaction \cite{Haldane1981,Haldane1981b}, which  allows 
	for stabilizing, for instance,  non Fermi liquid phases with, e.g., ``fractional'' tunneling of excitations with charge, but without spin,
	and vice versa \cite{Kane1992,Kane1992b}.
	Also, the onset of multiparticle scattering processes \cite{Nayak1997} gives rise to non Fermi liquid stable phases at strong enough values of the interaction in three-wire junctions, both in the spinless \cite{Oshikawa2006}, as well as in the spinful \cite{Hou2012}, case. 
	
	Stabilizing non-Fermi liquid phases typically requires un physically high values of the bulk interaction strength, which gives rise to relevant operators destabilizing the Fermi liquid phase(s) and, at the same time, stabilizing the non-Fermi liquid ones. Alternatively, at small values of the bulk interaction, or even 
	in the noninteracting limit, the onset of non Fermi liquid phases may be determined by the interaction between the collective modes 
	of the leads and local degrees of freedom emerging at the  junction.
	
	In particular, due to the high versatility of the Tomonaga-Luttinger liquid approach in describing
	one-dimensional spin chains in the spin-liquid phase \cite{Eggert1992}, as well as one-dimensional Josephson junction arrays \cite{Glazman1997,Giuliano2005}, or 
	pertinently engineered cold atom systems \cite{Cazalilla2011}, (Kondo-like) models of local magnetic impurities interacting with the collective 
	excitations of the leads  have been proposed in junctions of spin chains
	\cite{Tsvelik2013,Tsvelik2013b,Crampe2013,Giuliano2016a}, of one-dimensional Josephson junction networks \cite{Giuliano2013,Giuliano2009}, of cold atom condensates \cite{Buccheri2016}.
	At variance, local fermionic degrees of freedom  can emerge  as Klein factors, that is, real fermion operators required on implementing the bosonization over 
	a junction with more than two leads \cite{Oshikawa2006,Crampe2013}.
	Finally, localized degrees of freedom are clearly present when the leads are proximity-coupled at one end to a central island, i.e., a mesoscopic system, either grounded or floating, with a finite charging energy. The number of degrees of freedom is typically limited due to the physical size, as in quantum dots, or by a gapped spectrum, which suppresses the excitations at low temperatures. A remarkable example is provided by localized Majorana zero modes (MZMs) in junctions involving superconducting islands 
	\cite{Beri2012,Altland2013,Altland2014}.
	On entangling with the lead degrees of freedom, or with each other, Klein factors and MZMs trigger the onset of nontrivial phases and/or phase transitions in junctions of QWs, for many of which a full theoretical description is still lacking, or 
	applicable only to leading order in the distance (in parameter space) from other phases for which there is a complete theoretical model
	\cite{Oshikawa2006,Hou2012,Rahmani2012,Affleck2013,giuliano2019,Kane2020}. 
	
	An efficient mean to identify the system phases is by looking at the equilibrium charge and energy transport properties of the system under investigation (see, for instance, \cite{Pekola2021} for a review). These are typically not independent of each other: whenever an electronic system can be adiabatically deformed into a noninteracting Fermi gas, the ratio 
	between the charge and the thermal conductance of the system are related by 
	the Wiedemann-Franz law (WFL). This states that,  at temperatures   low with respect to the Fermi energy, such ratio is proportional to the temperature, through a universal constant  $L_0$, dubbed Lorenz number \cite{Franz1853}
	\beq
	L_0 = \frac{\pi^2 k_B^2}{3 e^2} 
	\approx 2.44 \times 10^{-8}W\Omega K^{-2}
	\:\:\:\: . 
	\label{lorenz_number}
	\eneq
	\noindent
	In practice, the WFL is experimentally well-verified in metals and semiconductors at room temperature \cite{Kumar1993}, though interaction effects and inelastic scattering may lead to a renormalization of the Lorenz number   \cite{Kane1996,Kane1997,Lavasani2019}.
	Also, localized interactions, e.g., such as in quantum dots, may provide an analogous renormalization of $L_0$ by a nonuniversal value \cite{Kubala2008,Karki2020c,Majidi2021}.
	
	It is worth mentioning that the WFL is expected to be satisfied when a one-dimensional system is connected to external Fermi liquid  
	reservoirs \cite{Ponomarenko1995,Safi1995,Maslov1995,Oshikawa2006}. In the same way, one does not expect deviations of the Lorenz number  
	from   $L_0$ in the presence of the Kondo effect, in the co-tunneling regime of an interacting quantum dot, due to the emergence of Nozier\`es Fermi liquid phase
	\cite{Nozieres1974,Boese2001,Costi2010}. In interacting quantum wires, the results are analogous \cite{Balachandran2012}.
	Somewhat surprisingly, the WFL appears to be valid even in a non Fermi liquid phase, such as for the overscreened Kondo fixed point
	\cite{Affleck1991f,Affleck1991,Affleck1993}.  As long as charge and heat are carried by the same excitations, regardless of
	whether they are Landau quasiparticles, or collective modes of the Tomonaga-Luttinger liquid leads, the Lorenz ratio remains unchanged \cite{Karki2020}.
	
	A phase (a fixed point) of the junction is uniquely identified by the local relations between the fields describing the collective excitations in the wires. 
	Typically, these are recovered by looking at the (equilibrium) transport properties of the system (electrical conductance for charged wires, spin conductance for
	spin liquids, {\it et cetera}) \cite{Oshikawa2006,Rahmani2010,Rahmani2012}.
	However, when comparing the theoretical predictions with the experimental results, in many cases of physical interest this procedure  may be tainted by  spurious 
	effects, that may spoil the interpretation of the measurements. Therefore, to unambiguously identify the  phases setting in at junctions of QWs, it becomes important 
	to make combined measurements of the charge transport properties of the system and of additional, pertinently chosen, quantities. 
	
	In this paper, we investigate the charge and the energy transport through  a  junction of $N>2$ interacting QWs. In doing so, we are  able to determine under which conditions and by means of which physical mechanisms the WFL is violated at the  junction. 
	Physically, this takes place when the charge-carrying excitations are separated from the heat-carrying ones. Consistently, we show that the WFL is expected to break down whenever Andreev reflection and/or crossed Andreev reflection processes take place at the junction, possibly in combination with the normal reflection and the normal transmission of particles.
	The coexistence of normal and Andreev reflection/transmission makes the net charge flowing in 
	one direction different from the net number of particles flowing in the same direction, 
	thus leading to a remarkable "charge-heat separation".
	Based on this observation, we go through an analysis of the charge and transport conductances 
	across a junction of $N$ QWs, both in the case of noninteracting and of interacting leads. 
	We find that the WFL is violated only if the charge is not conserved
	at the junction, which typically requires coupling the QWs to an underneath superconductor. 
	Typically, in this case, the Lorenz number   is rescaled by a nonuniversal factor, which
	depends on the microscopic scattering processes at the central island  \cite{Averin1995}.  
	Conversely, we show that the WFL breaks down also with charge conservation holding at the junction 
	when the low-energy modes of the leads are coupled to MZMs  localized at the central island.
	The interaction allows for stabilizing phases characterized by multiparticle normal and Andreev  
	reflection/transmission processes at the central island,  at a fixed, universal 
	and predictable renormalization of the Lorenz ratio.
	Beside being genuine non Fermi liquid phases, 
	due to the multiparticle scattering processes being  the only ones that survive in 
	the zero-temperature limit, these phases feature a remarkable charge-heat separation \cite{Benenti2017}, 
	with the consequent breakdown of the WFL, in an universal and predictable way. 
	
	Importantly, in order to recover the breakdown of the WFL in a junction of normal wires, such as the ones studied 
	in \cite{Oshikawa2006,Hou2012}, we need to have an extremely strong bulk interaction in the wire. Alternatively, 
	coupling the wires with emerging MZMs at the central island can effectively stabilize the breakdown of the WFL at relatively 
	small interactions, or even with noninteracting leads \cite{Us_short2021}. 
	The formalism we develop here, based on the Tomonaga-Luttinger liquid approach, allows us to make sharp predictions on the charge and thermal conductance tensors at the low-temperature fixed point (phase), as well as on their scaling dependence on the temperature in its vicinity. 
	As an example, we extend the calculation in \cite{Us_short2021} for a junction providing a
	realization of the topological Kondo model (TKM) \cite{Beri2012,Altland2013,Herviou2016}.
	
	Within our analysis, we also discuss in detail the modifications induced in the various phases 
	and, more in general, in the whole topology of the phase diagram, when the junction is connected 
	(as it is typical, in transport measurements) to outer, Fermi liquid reservoirs. When  the 
	reservoirs are attached at a (finite) distance $\ell$ from the junction, at low enough energies/long enough wavelengths, the system becomes sensitive  to their presence.
	Specifically, this implies that, when considering the equilibrium (dc) charge and thermal conductances, the effects related to the bulk interaction in the leads, including the renormalization of the Lorenz ratio,
	are eventually washed out by the presence of the effectively 
	noninteracting reservoirs \cite{Ponomarenko1995,Safi1995,Maslov1995,Oshikawa2006}. Often, in transport measurements, 
	one has to consider the effects of the reservoirs. Therefore,  we devote part of our work to disentangle the 
	violation of the WFL determined by the bulk interaction in the leads from the one genuinely due to 
	the onset of multiparticle (anomalous) scattering processes at the junction. In doing so, we also argue how the reservoirs affect the scaling of the corrections to the fixed point values of the conductances and of the Lorenz ratio and how they may modify the phase diagram itself.
	Based on the results of this paper, in \cite{Us_short2021} we conclude 
	that  the emergence of MZMs and their action in inducing the topological Kondo effect  becomes the effective mechanism triggering the (universal) breakdown of the WFL at the junction and that such phenomenon can be exploited for detection of the presence of MZMs at the junction.
	While standard charge transport experiments cannot presently distinguish MZMs from other effects unambiguously \cite{Nichele2017,Chen2019,Yu2021,Lutchyn2018,Flensberg2021}, our proposal provides a new experimental test, aimed at ruling out such ambiguities.
	
	The paper is organized as follows:
	\begin{itemize}
		\item In Section \ref{sec:Noninteracting}, we derive the charge conductance tensor and  the thermal  conductance tensor at a junction of $N$ one-dimensional, noninteracting, spinless QWs  and we evidence the importance of Andreev and/or crossed Andreev reflection at the junction, in order for the WFL to break down.
		
		\item In Section \ref{sec:Conductance_tensor}, we set up the formalism to compute the charge and the thermal conductance tensor within the Tomonaga-Luttinger liquid approach. In particular, we trace out a direct correspondence between the conformal boundary conditions describing 
		a fixed point in the phase diagram of the junction and the corresponding charge and heat conductance tensors. 
		
		\item In Section \ref{sec:Examples}, we compute the charge and the thermal conductance tensor at various fixed points of a generic $N$-wire junction. We evidence the violations of the WFL and discuss the two mechanisms that originate it.
		
		\item In Section \ref{sec_Renormalization_3}, we discuss the WFL  in a junction of $N=3$ QWs \cite{Chamon2003,Oshikawa2006} and in the TKM \cite{Beri2012,Altland2013,Herviou2016}. 
		In reviewing the phase diagram of both systems, we also derive the scaling properties (with the temperature as scaling parameter) of the conductance tensors and of the Lorenz ratio close to each fixed point.  We analyze both the disconnected case and the one in which the junction is connected to external  reservoirs.
		
		\item In Section \ref{sec:Conclusions} we provide our conclusions and discuss about possible further developments of our work.
		
		\item In the various Appendices we provide mathematical details of our work.
		
	\end{itemize}

	To  help following the various abbreviations, we list in table 
	\ref{abbr} the meaning of  the ones we use most commonly throughout the paper.
	\begin{table}
		\centering
		\begin{tabular}{| c | c |}
			\hline 
			QW  & Quantum wire \\
			\hline 
			MZM & Majorana zero mode \\
			\hline
			WFL & Wiedemann-Franz law  \\
			\hline
			TKM & Topological Kondo model \\
			\hline
			CCT & Charge conductance tensor \\
			\hline
			HCT & Heat conductance tensor  \\
			\hline
			(D)FP & (Disconnected) Fixed point \\
			\hline
			RG & Renormalization group \\
			\hline
		\end{tabular}
		\caption{Glossary of most commonly used abbreviations} 
		\label{abbr}
	\end{table}
	\noindent

	\section{Charge and thermal conductance tensor at a junction of noninteracting quantum wires}
	\label{sec:Noninteracting}
	
	We now derive the charge conductance tensor (CCT) and the heat conductance tensor (HCT) at a junction 
	of $N$ one-dimensional, noninteracting, spinless  QWs, connected to each other at $x=0$.
	In the following we limit our analysis to the case in which the wires are all equal to each other, though this implies no loss
	of generality in our derivation. 
	As we focus on the equilibrium transport properties of the junction, 
	we resort to a low-energy, long-wavelength expansion around the Fermi points, 
	so to write the junction Hamiltonian as $H = H_{{\rm Fer} , 0} + H_B$. With $H_B$ we denote  the 
	boundary Hamiltonian encoding the system dynamics at the junction, while the ``lead'' Hamiltonian $H_{{\rm Fer} , 0} $ is given by 
	\beq
	H_{{\rm Fer} , 0} = - i v \: \sum_{ j = 1}^N \: \int_0^\ell \: d x \: \{ \psi_{R , j}^\dagger \partial_x \psi_{R , j }
	- \psi_{L , j}^\dagger\partial_x \psi_{L , j }  \} 
	\:\:\:\: . 
	\label{noni.1}
	\eneq
	\noindent
	In Eq.\eqref{noni.1} we denote with $\psi_{R/L , j }$  the chiral fermionic fields corresponding to the two chiral excitation  branches around the Fermi points, with the subscripts $R$ ($L$) labeling right-handed (left-handed) branches and $v$  the Fermi velocity. Also, we introduce the lead length $\ell$ as a large-distance regularization.
	Eventually, when computing physical quantities, we take the $\ell \to \infty$ limit. 
	\begin{figure}
		\center
		\includegraphics*[width=.8 \columnwidth]{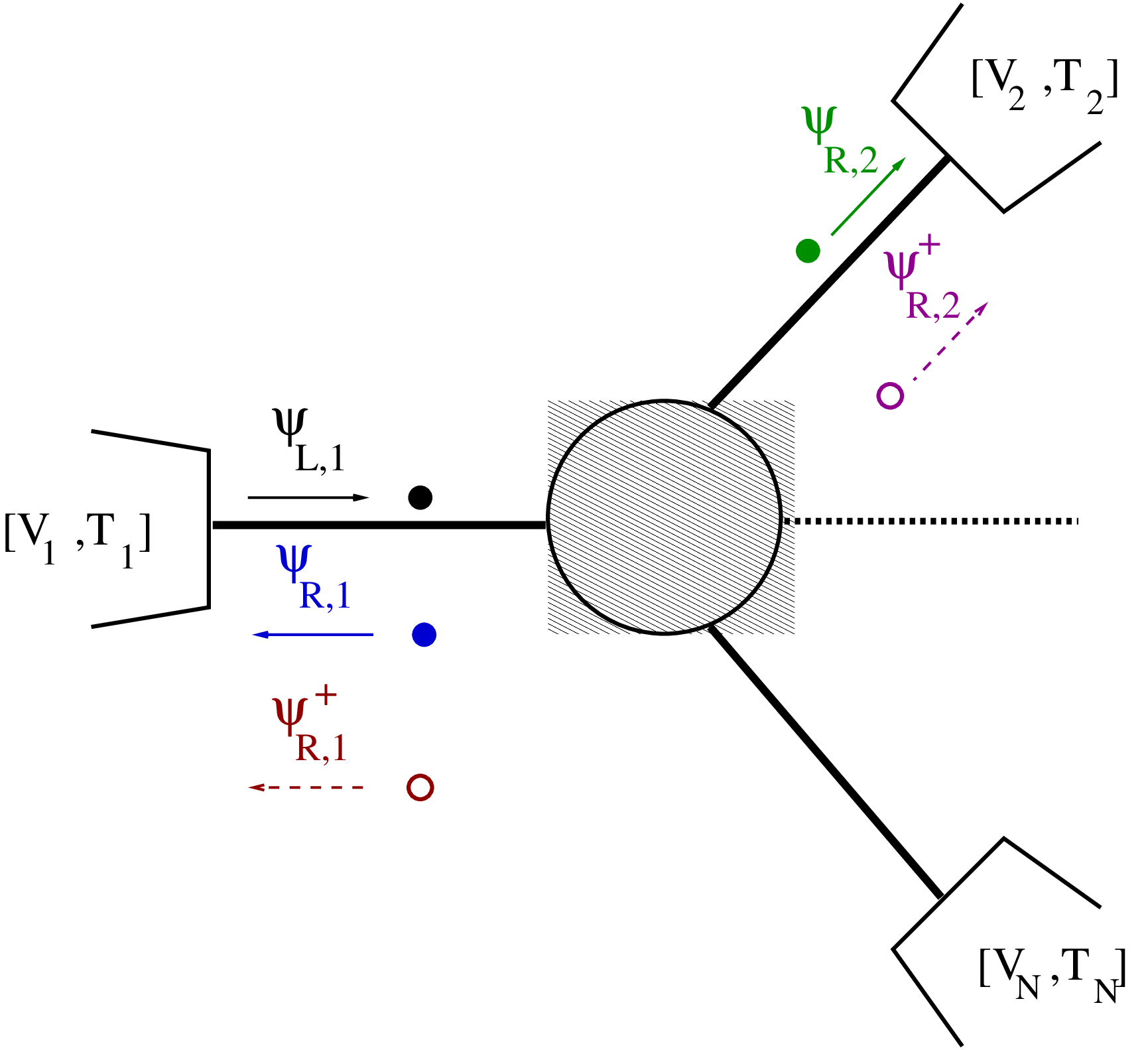}
		\caption{Sketch of the fermionic $N$ wire junction. Chiral fermions are shot toward the central island
			from the external reservoirs, biased at voltage and temperature $V_j , T_j$, with $j = 1 , \ldots , N$. 
			An incoming  $L$ electron from wire $j'$(=1 in the figure) (black dot) can be reflected within the same wire as an electron (``normal reflection'', 
			blue dot),  or reflected within the same wire as a hole (``Andreev reflection'', red empty  dot), or transmitted to 
			wire $j  (\neq j'$,=2 in the figure)  as an electron (``normal transmission'', green dot), or, finally, transmitted to wire $j$ as 
			a hole (``crossed Andreev reflection'', magenta empty dot).} 
		\label{extended_junction}
	\end{figure}
	
	To encompass also the case of junctions involving superconductors, we consider all the possible 
	single-particle scattering processes, which we draw in Fig.\ref{extended_junction}. Specifically, we see that  an incoming 
	$L$ electron from wire $j$  can be reflected within the same wire as an electron (``normal reflection'' ),  or reflected
	within the same wire as a hole (``Andreev reflection''), or transmitted to 
	wire $j' (\neq j)$ as an electron (``normal transmission''), or, finally, transmitted to wire $j'$ as 
	a hole (``crossed Andreev reflection''). Denoting respectively with $r_{j',j'} , a_{j',j'} , t_{j',j}, 
	c_{j',j}$ the corresponding scattering amplitudes, we encode them in the ``extended'' $2N \times 2N$ ${\bf S}$ matrix 
	\begin{eqnarray}\label{noni.2}
		\psi_{R , j' } ( 0 ) &=& \sum_{j = 1}^N \{ S_{j',j} \psi_{L,j} ( 0 ) + S_{j',j+N} \psi_{L , j}^\dagger ( 0 ) \}
		\\
		\psi_{R , j'}^\dagger ( 0 ) &=& \sum_{ j = 1}^N \{ S_{j'+N,j} \psi_{L , j} ( 0 ) + S_{j'+N,j+N} \psi_{L , j}^\dagger ( 0 ) \}\nonumber
		\: . 
	\end{eqnarray}
	\noindent
	In Eq.(\ref{noni.2}) we have labeled the ($2N \times 2N$) ${\bf S}$ matrix so that, in the 
	matrix elements $S_{a,b}$,   indices $a (b) = 1 , \ldots , N$ refer to particles, indices $a (b) = N + 1 , \ldots , 2 N$
	refer to holes. Assuming, as we do throughout all our paper, that particle-hole symmetry holds in our system, we infer that the ${\bf S}$ matrix must satisfy the Bogoliubov-de Gennes constraint \cite{Kane2020},
	${\bf S}^\dagger= K_C {\bf S} K_C$, with $K_C = \sigma^x \otimes {\bf I}_N$, where the Pauli matrix $\sigma^x$ acts in
	particle-hole space. This implies
	(assuming $j' = 1, \ldots , N$)
	\begin{eqnarray}
		S_{j',j} &=& \delta_{j',j} r_{j',j'} + [ 1 - \delta_{j',j} ] t_{j',j} \nonumber \\
		S_{j',j+N} &=& \delta_{j',j} a_{j',j'} + [ 1 - \delta_{j',j} ] c_{j',j} \nonumber \\
		S_{j'+N,j} &=& S_{j',j+N}^* \nonumber \\
		S_{j'+N,j+N} &=& S_{j',j}^*
		\:\:\:\: . 
		\label{noni.3}
	\end{eqnarray}
	\noindent
	Going through a similar derivation, we recover the ${\bf S}$-matrix encoding the scattering of an incoming 
	hole from lead $j$ throughout the wires connected to each other at the central island  (see Appendix \ref{Noninteracting} for
	details).  Keeping only the linearly dispersing low-energy modes, the 
	${\bf S}$ matrix elements in Eqs.\eqref{noni.2} become independent of the energy. While such an 
	approximation does not affect our following derivation, we refer to Appendix \ref{Noninteracting} for a full discussion of the scattering amplitudes within a lattice fermionic model for the leads, which allows to retain the full energy dependence of the ${\bf S}$ matrix at any step of the derivation of the
	conductance tensors.
	
	Various proposals have recently been formulated, about 
	realizing junctions exhibiting either Andreev, or crossed Andreev reflection, or both. For instance, it has been pointed out that 
	Andreev reflection and crossed Andreev reflection can become relevant processes in junctions realized by depositing quantum wires on top of 
	a superconducting island with finite charging energy $E_c$ and Josephson coupling to a  superconductor \cite{Eriksson2014,Eriksson2014_PRL,Beri2012,Herviou2016,Altland2013}. In this case, Andreev reflection and crossed Andreev reflection are triggered by the emergence of localized MZMs 
	\cite{Kitaev2001,Beri2012,Altland2013}, thus giving rise to a remarkable ``topological'' Kondo effect. A similar physics is expected to arise at a junction between several quantum wires, or a multichannel quantum wire, and a topological superconductor \cite{Affleck2013,giuliano2019,Kane2020}.
	More generically, Andreev reflection and/or crossed Andreev reflection can arise at junctions of normal quantum wires, but at the cost of having a strong electronic interaction within each wire \cite{Oshikawa2006,Hou2012}. Finally, we note that nonlocal crossed Andreev reflection can be in principle recovered across a finite-size one-dimensional 
	topological superconductor with long range pairing and/or electron hopping \cite{Nilsson2008,Giuliano2018_2}.
	
	Here, given the extended ${\bf S}$ matrix at the junction, we now compute the 
	equilibrium CCT and HCT 
	of a junction connected to external reservoirs, such as the one we sketch in Fig.\ref{extended_junction}.
	We regard the external reservoirs as noninteracting Fermi liquids, each one characterized by a voltage bias $V_j$ and by a
	temperature $T_j$. Paraphrasing \cite{Anantram1996}, we describe each of them by means of the  Fermi distribution functions 
	for a single particle-like eigenmode 
	at energy $\epsilon$,  $f^{(p)}_j ( \epsilon )$, and for a hole-like eigenmode at energy $\epsilon$, $f^{(h)}_j ( \epsilon )$.
	The distribution  functions are respectively given by 
	
	\begin{eqnarray}
		f^{(p)}_j ( \epsilon ) &=& \frac{1}{1 + e^{ \beta_j ( \epsilon - \mu - e V_j)}} \nonumber \\
		f^{(h)}_j ( \epsilon ) &=& \frac{1}{1 + e^{ \beta_j ( \epsilon - \mu + e V_j)}}
		\;\;\;\; , 
		\label{noni.4}
	\end{eqnarray}
	\noindent
	with $\mu$ being the common reference (for all the leads) chemical potential and $\beta_j = [ k_B ( T + \delta T_j )]^{-1}$, 
	with $T$ being the common reference temperature. Of course, in order for linear response theory (which we employ in 
	the following) to apply, we require that $ | e V_j / \mu | , | \delta T_j / T | \ll 1$, $\forall j = 1 , \ldots , N$. 
	In order to compute the CCT and the HCT, in the following we look at the average values of 
	the electric current density  operator in lead $j$, $j_{{\rm el} , j }$ and of the {\it energy}
	current density operator  in lead $j$,
	$j_{{\rm th} , j}$. These can be readily recovered by means of the appropriate continuity 
	equations for the electric charge density operator and for the energy density operator. They are 
	given by 
	
	\begin{eqnarray}
		j_{{\rm el} , j } ( x , t ) &=& e v \: \left\{ \psi_{R ,j }^\dagger  \psi_{R , j }  - 
		\psi_{L , j }^\dagger  \psi_{L , j }  \right\} \nonumber \\
		j_{{\rm th} , j } ( x , t ) &=& -  i v^2  \: \left\{  \psi^\dagger_{R , j }  \partial_x \psi_{R , j }  + \psi_{L , j }^\dagger \partial_x \psi_{L , j } \right\}\:\: . 
		\label{noni.4.1}
	\end{eqnarray}
	\noindent
	Denoting with $I_{{\rm el} , j}$
	and with $I_{{\rm th} , j }$ the expectation values of the operators in Eqs.\eqref{noni.4.1}, as detailed in Appendix \ref{Noninteracting},
	we obtain the CCT and the HCT  matrix elements  
	\begin{eqnarray}
		G_{j,j'} &=& \frac{e^2}{2 \pi} \left(- \delta_{j,j'} + T_{j,j'} - A_{j,j'} \right) \nonumber \\
		K_{j,j'} &=& \frac{\pi k_B^2 T}{6} \: \left( -\delta_{j,j'}  + T_{j,j'} + A_{j,j'} \right)
		\:\: ,
		\label{noni.6}
	\end{eqnarray}
	\noindent
	with $T_{j,j'}=| t_{j , j'} |^2$ if $j\ne j'$, while $ T_{j,j}=| r_{j} |^2$, as well as $A_{j,j'} = | c_{j,j'} |^2$ if $j\ne j'$, 
	while $A_{j,j}=| a_{j} |^2$. The thermal current obeys the Kirchhoff law \cite{Benenti2017}, from which
	\beq
	\sum_{ j' = 1}^N \{ T_{j ,j'} + A_{j ,j'} \} = \sum_{j' = 1}^N \{ T_{j',j } + A_{j',j } \} =1\: ,
	\label{noni.8}
	\eneq
	\noindent
	$\forall j  = 1 , \ldots , N$. This is a direct consequence of the unitarity of the extended ${\bf S}$-matrix.
	
	If the charge is conserved at the junction, then Kirchhoff laws holds for the electric current as well and \mbox{$A_{j',j}=0$}, $\forall j,j'$ 
	necessarily.
	In a junction of multiple QWs, we modify the temperature or the electrochemical potential in one of the reservoirs and measure the charge or heat current that exits the junction in another wire. Therefore, a natural extension of the definition of Lorenz ratio in our geometry is  \cite{Benenti2017}
	\beq
	\mathcal{L}_{j,j'} = \frac{K_{j,j'}}{T G_{j,j'} } 
	\;\; , \label{noni.x}
	\eneq
	and using Eqs.\eqref{noni.6}, we readily obtain the WFL 
	\mbox{$
		K_{j,j'} 
		= T L_0 G_{j,j'}$},  with $L_0$ in Eq.\eqref{lorenz_number}.
	Eq.\eqref{noni.x} is defined only when the denominator is non-zero. In the opposite case, for $j,j'$ such that $G_{j,j'} = 0$, the correct result is obtained via perturbation theory, which will be addressed in sections \ref{sec:Conductance_tensor} and \ref{sec_Renormalization_3}. 
	Instead, if the junction does not conserve the electric charge, as in the presence of a grounded superconductor, we do not expect the Kirchhoff law for the electric current to hold. We therefore obtain
	\beq
	\mathcal{L}_{j,j'} =
	\: \frac{T_{j,j'} + A_{j,j'}-\delta_{j,j'}}{T_{j,j'} - A_{j,j'}-\delta_{j,j'}}  L_0 
	\:\:, 
	\label{noni.12}
	\eneq
	Let us consider Eq.\eqref{noni.12} for $j\ne j'$: we see that, while the contributions to the 
	thermal conductance by normal transmission and crossed Andreev reflection have the same sign, the ones to the electric conductance have opposite sign, which, in general, implies a  Lorenz ratio 
	${\cal L} > L_0$ as soon as $A_{j,j'} > 0$. This is a consequence of the fact 
	that outgoing particles and holes move in the same directions, but with opposite charges.
	Therefore, while the energy currents carried by the two of them add up, the electric currents 
	get subtracted from each other. Similarly, considering the ratios between 
	the diagonal conductances in Eq.\eqref{noni.12}, we see that  normal reflection  and Andreev reflection both lower 
	the thermal current within lead $j$, while the former lowers and the latter increases the electric current.
	In conclusion,  we have established that, whenever only single-particle  
	scattering processes take place at the central island (as it typically happens when, e.g., 
	the leads are noninteracting) and if the total charge is conserved, no violation of the WFL can be realized at 
	a junction of $N$ quantum wires. This conclusion can be circumvented by having a nonzero interaction within the leads and/or
	by physically relevant mechanisms, which can stabilize nontrivial phases of the system characterized by multi-particle scattering processes at the junction. 
	In order to study a number of system in which this takes place, we now generalize our derivation to the case of interacting 
	leads, which requires to resort to the  framework of Abelian bosonization for one-dimensional systems.

	\section{Transport at a junction of interacting quantum wires and conformal boundary conditions}
	\label{sec:Conductance_tensor}
	
	We now set up the formalism to compute the charge and the thermal conductance tensor at a junction of interacting ballistic QWs  
	within the Luttinger liquid approach of \cite{Kane1992,Kane1992b,Kane1996,Kane1997}. 
	On one hand, this allows us to account for a nonzero bulk interaction in the leads,
	on the other hand, it enables us to generate boundary conditions different from the 
	one arising from single-particle scattering discussed in Section \ref{sec:Noninteracting}. 
	
	\subsection{Electric and thermal conductance}
	Within the Tomonaga-Luttinger liquid approach, the chiral fermionic operators in each lead are realized as functionals of  
	the plasmon fields $\phi_j ( x )$,
	together with their conjugates $\theta_j ( x )$,   satisfying  the  commutation relations 
	$[\phi_j ( x ) , \theta_{j'} ( x' ) ]  = \frac{i}{2}  \delta_{j,j'} \epsilon (x-x')$, with $\epsilon (x-x')$ being 
	the sign function, with all the other commutators being zero.  Specifically, they are represented as the vertex operators
	\begin{eqnarray}
		\psi_{R , j } ( x ) &=& \Gamma_j \: e^{ i \sqrt{\pi} [ \phi_j ( x ) + \theta_j ( x ) ]} \nonumber \\
		\psi_{L , j } ( x ) &=& \Gamma_j \: e^{ i \sqrt{\pi} [ \phi_j ( x ) - \theta_j ( x ) ] } 
		\;\;\;\; , 
		\label{eq.3bis}
	\end{eqnarray}
	\noindent
	with the Klein factors $\Gamma_j$ being real fermion operators, satisfying the 
	algebra $\{ \Gamma_j , \Gamma_{j'} \} = 2 \delta_{j,j'}$, introduced in order to
	assure the appropriate anticommutation relations between single electron annihilation/creation 
	operators acting over different leads. The dynamics of the lead bosonic fields is encoded 
	in the   Hamiltonian $H_{ 0 , {\rm Bos}}$, given by  
	
	\beq
	H_{0 , {\rm Bos}} = \frac{u}{2} \: \sum_{ j = 1}^N \: \int_0^\ell \: d x \: 
	\left[ g \left( \partial_x \phi_j ( x ) \right)^2 + g^{-1} \left( 
	\partial_x \theta_j ( x ) \right)^2 \right]
	\:\:\:\: , 
	\label{eq.4}
	\eneq
	\noindent
	with $u$ being the collective plasmon velocity and $g$ being the dimensionless 
	Luttinger parameter (again, for the sake of simplicity, we assume that all the wires are characterized by the same 
	parameters $g$ and  $u$. Yet, our derivation can be readily extended to a junction of Luttinger liquids  with different parameters
	by a pertinent implementation of, e.g., the methods developed in \cite{Hou2012,Affleck2013}), 
	complemented with the appropriate boundary conditions at the ``inner'' boundary 
	$x = 0$.  In terms of the fields $\phi_j $ and $\theta_j$, the charge density and the charge  current  operators in lead $j$, 
	$\rho_{{\rm el} , j } ( x , t ) , j_{{\rm el} , j } ( x , t )$, are given by 
	
	\begin{eqnarray}
		\rho_{{\rm el} , j } ( x , t ) &=& \frac{ e }{\sqrt{\pi}} \partial_x \phi_j ( x , t ) \nonumber \\
		j_{{\rm el} , j } ( x , t ) &=& \frac{ e    u g}{\sqrt{\pi}} \partial_x \theta_j ( x , t )
		\;\;\;\; , 
		\label{spec.1}
	\end{eqnarray}
	\noindent
	and are related to each other via a continuity equation. In the same way, the energy current $j_{{\rm en} , j }$ can be defined from the continuity equation for the Hamiltonian density in \eqref{eq.4} as
	\begin{eqnarray}
		j_{{\rm en} , j } ( x , t ) &=& u^2 \partial_x \phi_j ( x , t ) \partial_x \theta_j ( x, t ) 
		\;\;\;\; . 
		\label{spec.2}
	\end{eqnarray}
	\noindent 
	The heat current $j_{{\rm th} , j }$ follows from the above expression as $j_{{\rm th} , j }=j_{{\rm en} , j }-V_j j_{{\rm el} , j }$ \cite{Benenti2017}. In fact, at  charge neutrality, the energy and thermal currents yield the same results for the conductances in the linear response regime.
	
	Working with ballistic QWs, in the following we will see that the 
	CCT and the HCT  are only affected by 
	the  scattering processes  at the central island.
	While our approach is effective in working out the zero-temperature, 
	fixed point properties of the junction,   in general  other 
	effects, which we do not consider here, such as coupling with phonons, may become effective in determining the 
	finite-temperature transport properties of our system \cite{Chernyshev2016} (see, e.g.,  \cite{Lucas2018,Lavasani2019}
	for a comprehensive discussion of several possible physical mechanisms affecting 
	the thermal transport properties of an 
	electronic system). 
	
	\begin{figure} 
		\center
		\includegraphics*[width=0.8 \columnwidth]{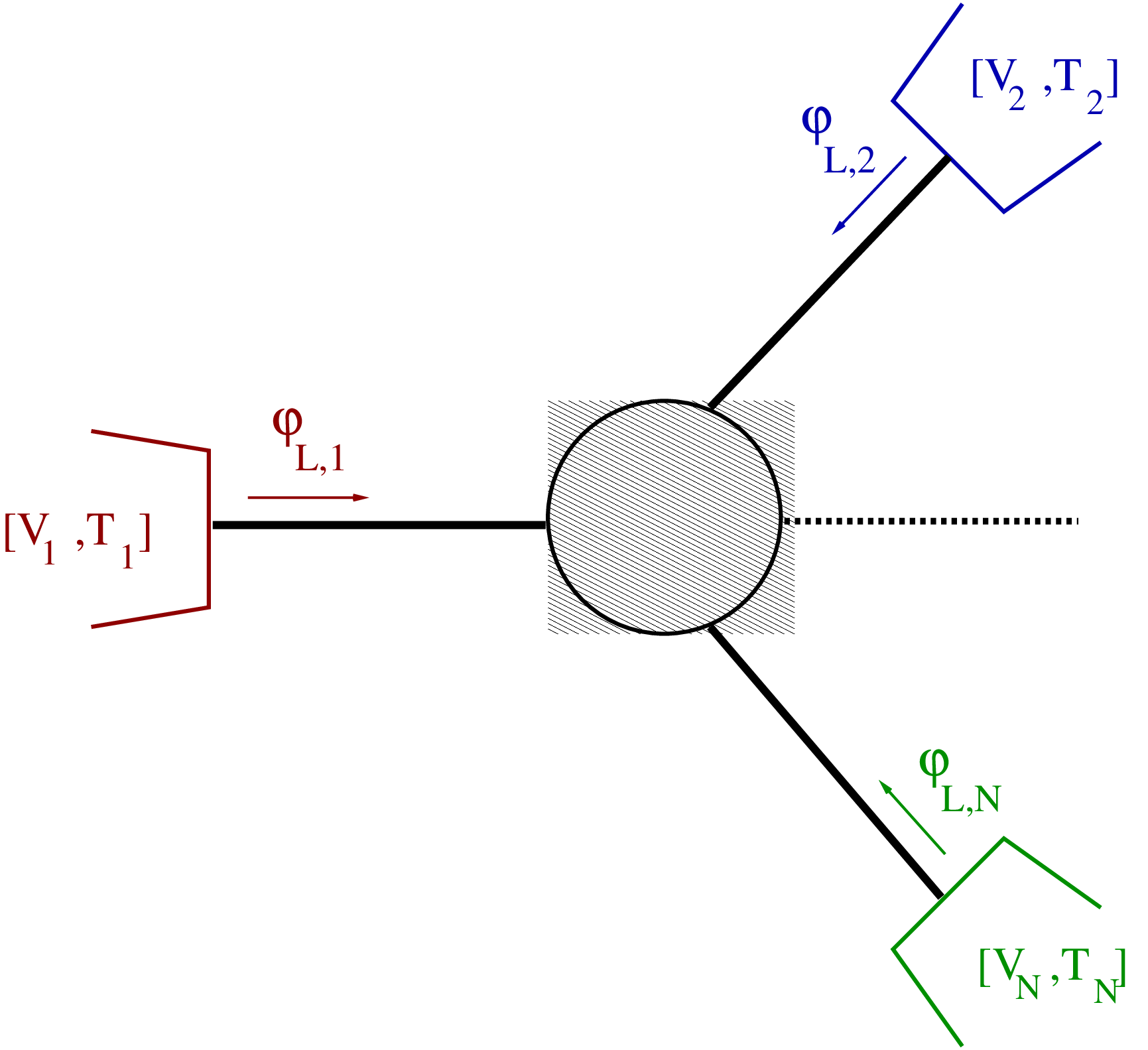}
		\caption{Sketch of a junction of $N$ interacting quantum wires. Each wire $j$ is connected to a   reservoir, which 
			injects into the system left-handed modes at voltage bias $V_j$ and at temperature $T_j$. The dashed 
			region represents the central island, whose dynamics is encoded in the matrix ${\bf \rho}$, relating the right-handed to 
			the left-handed chiral modes at $x=0$. } 
		\label{junction}
	\end{figure}
	\noindent
	In Fig.\ref{junction} we provide a sketch of our junction: within a  generalization of the calculation for a single
	wire discussed in  \cite{Kane1996,Kane1997},
	the QW $j$ is connected to an external reservoir that injects chiral, left-handed modes, at chemical potential $V_j$ and temperature $T_j$. Accordingly,   we introduce the chiral bosonic fields 
	in the lead $j$,   $\varphi_{R , j} ( x ) , \varphi_{ L ,j} ( x )$: they are related 
	to the fields $\phi_j ( x ) , \theta_j ( x )$   in Eq.\eqref{eq.4} via the relations 
	\begin{eqnarray}
		\phi_j ( x ) &=& \frac{\varphi_{R , j } ( x ) + \varphi_{L , j } ( x ) }{\sqrt{g}} \nonumber \\
		\theta_j ( x ) &=& \sqrt{g} \{ \varphi_{R , j } ( x ) - \varphi_{ L , j} ( x ) \} 
		\:\:\:\: . 
		\label{junc.1}
	\end{eqnarray}
	We describe the junction by means of pertinent conformal boundary conditions  between the bosonic fields: 
	in particular, the relation between the $R$ and the $L$ fields is encoded in 
	the $N\times N$ splitting matrix ${\bf \rho}$ \cite{Oshikawa2006,Bellazzini2008,Bellazzini2009}
	\begin{equation}
		\varphi_{ R , j} ( 0 ) = \sum_{ j' = 1}^N   \rho_{ j , j'}   \varphi_{ L , j' } ( 0 ) 
		\:\: . 
		\label{eq:rho}
	\end{equation}
	By requiring that Eq.(\ref{eq:rho}) is consistent with 
	the canonical commutation relations between 
	the $\varphi_{ R , j}$ and the $\varphi_{ L , j}$ fields, one readily finds that 
	${\bf \rho}$ must be an orthogonal matrix.   Typically, relations such 
	as the ones in Eq.\eqref{eq:rho} hold at a conformally invariant fixed point of the phase diagram
	of the junction, where scale  invariance implies that the splitting matrix does not depend on  the momenta.  
	Using Eq.\eqref{eq:rho}, we  define $N$ ``unfolded'' fields 
	\beq
	\varphi_j ( x ) = \Biggl\{ \begin{array}{l}
		\varphi_{L, j }  ( x ) \;\;\; , \;\; (0 \leq x \leq \ell ) \\
		\sum_{ j' = 1}^N \: \rho_{ j' , j} \varphi_{R , j'}  ( - x ) \;\;\; , \;\;
		(-\ell \leq x < 0)
	\end{array}
	\:\:\:\: . 
	\label{6}
	\eneq
	\noindent
	By construction, each field $\varphi_j ( x )$ is at chemical and thermal equilibrium with 
	the reservoir at voltage bias $V_j$ and at temperature $T_j$. Accordingly, at nonzero 
	biases, we rewrite the lead Hamiltonian in 
	terms of the unfolded fields as $H_{0 , {\rm Bos}} = \sum_{j = 1}^N H_{0 , {\rm Bos} , j}$,
	with
	\begin{eqnarray}
		H_{0 , {\rm Bos} , j} &=&   \int_{-\ell}^\ell dx \left\{ u( \partial_x \varphi_j ( x ))^2 
		+ e \sqrt{\frac{g}{\pi}} V_j \partial_x \varphi_j ( x ) \right\} 
		\:\:\:\: . 
		\label{junct.5}
	\end{eqnarray}
	\noindent
	Thus, once a generic observable ${\cal O}$ is expressed in terms of the fields $\varphi_j$, we 
	compute its thermal average as 
	
	\beq
	\langle {\cal O} \rangle = \frac{{\rm Tr} [ {\cal O} e^{ - \sum_{j  = 1}^N \beta_j H_{0, {\rm Bos} , j } } ] }{\prod_{j = 1}^N {\cal Z}_j 
		[ V_j , \beta_j ]  } 
	\:\:\:\: , 
	\label{junct.6}
	\eneq
	\noindent
	with  $\beta_j  = (k_B T_j)^{-1}$ and
	\beq
	{\cal Z}_j [ V_j , \beta_j ] = {\rm Tr} [ e^{ - \beta_j H_{0,{\rm Bos} , j}  } ]
	\;\;\;\; . 
	\label{junct.7}
	\eneq
	\noindent
	In terms of the chiral fields  the   electric and thermal current operators in lead $j$ are
	\begin{eqnarray}
		j_{{\rm el} , j } ( x  , t ) &= &e u \sqrt{\frac{g}{\pi}} \: \left[ 
		\partial_x\varphi_j (u t_+) - \sum_{ j' = 1}^N \rho_{ j' , j } \partial_x \varphi_{j'} ( u t_-) \right] 
		\nonumber\\ \label{junct.8} 
		j_{{\rm th} , j} ( x , t ) &=& - \left[u\partial_x\varphi_j  (u t_+ )\right]^2  +
		\left[u\sum_{ j' = 1}^N \rho_{j' , j} 
		\partial_x\varphi_{j'} (ut_-) \right]^2 
		\:,
	\end{eqnarray}
	where $t_\pm = t\pm x/u$. The effect of the potential bias can be reabsorbed in a shift of the fields
	\beq
	\partial_x \bar{\varphi}_j ( t_\pm  )  = \partial_x \varphi_j ( t_\pm  ) \pm  \frac{e}{2 u} \sqrt{\frac{g}{\pi}} V_j 
	\:\:\:\: . 
	\label{junct.9}
	\eneq
	\noindent
	Switching to the shifted fields in Eq.(\ref{junct.9}), it is now straightforward to 
	implement the formalism of Appendix \ref{ballistic_single}, pertinently generalized to an $N$ QW junction, to 
	compute the average values of the current operators.   Retaining only linear contributions in the applied biases, 
	we eventually obtain the electric and thermal conductance tensors,
	
	\begin{eqnarray}
		G_{j , j'}  &=& \frac{ e^2 g}{2 \pi} \: \{\rho_{j, j'} -   \delta_{ j , j'} \} \label{junct.12G} \\
		{ K}_{ j  , j '}  &=& \frac{\pi  k_B^2 T}{6} \: \{ \rho_{j , j'}^2 - \delta_{ j , j'}  \}
		\;.
		\label{junct.12K}
	\end{eqnarray}
	In Eqs.(\ref{junct.12G},\ref{junct.12K}) we denote with $T$  the equilibrium, reference temperature of the reservoirs and, by 
	definition, we assume that  the currents  exiting  the central island always have
	positive sign.  Similar equations have been derived in Ref.\cite{Mintchev2013} in the framework of a 
	Luttinger liquid in a nonequilibrium steady state.   The orthogonality of ${\bf \rho}$ readily implies the Kirchhoff law for the thermal conductance tensor, $\sum_{ j = 1}^N K_{j,j'} = \sum_{j' = 1}^N K_{j , j'} = 0$.

	Concerning the results in Eqs.\eqref{junct.12G} and \eqref{junct.12K}, it is worth stressing that, throughout the paper, we always 
	assume  that particle-hole symmetry holds at equilibrium, which implies that the Seebeck 
	and Peltier coefficients vanish. Formally, this can be traced back to the $\mathbb{Z}_2$ symmetry of the bosonic Hamiltonian 
	Eq.\eqref{junct.6}, for $V_j=0$.
	More generally, in the context of a Tomonaga-Luttinger liquid,  particle-hole symmetry breaking may either be determined by, e.g., bulk cubic (or higher-order) interactions arising from nonlinear terms in the fermion dispersion relations, or by local, Sine-Gordon like interactions \cite{Kane1996}.  In the former case, symmetry breaking operators are typically infrared irrelevant
	and can be safely neglected throughout our derivation . The latter case takes place for energy-dependent (bare) boundary interaction strengths, a situation not considered in this work. 
	
	Taking the ratio between the thermal conductance and the electric conductance across any two leads, one has the Lorenz 
	ratio
	\begin{equation}
		\mathcal{L}_{j , j'} = \frac{L_0}{g}
		\left(  \rho_{j , j'} + \delta_{j , j' } \right) 
		\; .
		\label{junct.13}
	\end{equation}
	In writing Eq.\eqref{junct.13}, it is implied that $G_{j,j'}\ne0$. Instead, when $G_{j,j'}=0$ (and consequently $K_{j,j'}=0$), as it happens, for instance, at the disconnected fixed point of a junction of $N$ wires discussed in the following, the ratio has to be computed using the finite-$T$ corrections to  the conductances within the framework of Appendix \ref{General:expression}. 
	
	From  Eq.\eqref{junct.13} we identify the two factors that renormalize the Lorenz ratio: a contribution stemming from the interaction in the QWs, encoded in the Luttinger parameter $g$ \cite{Kane1992}, and a term determined by the tensor structure dictated by the splitting matrix. The former contribution  is washed out when the junction is connected to Fermi liquid  
	reservoirs \cite{Ponomarenko1995,Safi1995,Maslov1995,Oshikawa2006} (see Appendix \ref{Renormalization} for an extensive discussion about this 
	point), while the  latter term can lead to a violation of the WFL even in the absence of interactions in the QWs.
	As we are interested in violations of the WFL stemming from the dynamics at the junction, in all the examples that we discuss 
	in the following we attempt to disentangle the two effects and focus onto the contribution arising from   the splitting matrix 
	corresponding to a given  fixed point.
	
	\subsection{The Wiedemann-Franz law in the $N=3$-junction.}
	\label{genrho_3}
	
	We now perform a comprehensive analysis of the (violation of) the WFL at the fixed points of a $N=3$ junction of QWs. To 
	do so, we next review the general parametrization of the ${\bf \rho}$ matrices 
	describing conformal boundary conditions  in an $N=3$ junction in bosonic coordinates. Specifically, we first resort to  a purely algebraic classification, 
	without  addressing the issue of the stability of a specific fixed point (FP)  {\cite{Oshikawa2006,Bellazzini2006,Bellazzini2008,Bellazzini2009}. At a second stage, we discuss
		the phase diagram and the FPs that describe the low-temperature physics of specific systems as particular cases of the general results. 
		It is also important to point out that some FPs of the $N=3$ junction, such as the {\bf M}-fixed point of 
		\cite{Oshikawa2006}, cannot  be described in terms of simple conformal boundary conditions, 
		in bosonic coordinates. However, they admit (in the presence 
		of Fermi liquid  reservoirs) a  description in terms of a fermionic scattering matrix \cite{Giuliano2015}, which, as discussed in section 
		\ref{sec:Noninteracting}, implies that the WFL is automatically satisfied.
		
		Let us first assume total charge conservation, i.e., that the electric current in Eq.\eqref{junct.8} satisfies the Kirchhoff's law at the junction, 
		as well as the invariance under the $\mathbb{Z}_3$ transformation exchanging the leads with each other.  Requiring that ${\bf \rho}$ is orthogonal, as it must be in order to preserve the canonical commutation relations between the bosonic fields, the splitting matrix depends only on the Luttinger parameter $g$ and a real parameter $\vartheta$ \cite{Oshikawa2006} as
		\beq
		{\bf \rho} ( \vartheta )  =
		\left( \begin{array}{ccc}
			a ( \vartheta ) & b ( \vartheta )     & c ( \vartheta )   \\
			c ( \vartheta ) & a ( \vartheta )     & b ( \vartheta )   \\
			b ( \vartheta ) & c ( \vartheta )     & a ( \vartheta ) 
		\end{array} \right)
		\;\; , 
		\label{para.3}
		\eneq
		with  $-\pi<\vartheta\le\pi$ and 
		\begin{eqnarray}
			a ( \vartheta ) &=&  \frac{3 g^2 - 1 + ( 3 g^2 + 1 ) \cos ( \vartheta) }{ 3 [ 1 + g^2 + ( g^2 - 1 ) \cos ( 
				\vartheta) ] }  \nonumber \\
			b ( \vartheta ) &=& \frac{2 [ 1 - \cos ( \vartheta ) + \sqrt{3} g \sin ( \vartheta) ]  }{ 3 [ 1 + g^2 + ( g^2 - 1 ) \cos ( 
				\vartheta) ] }  \nonumber \\
			c ( \vartheta ) &=& \frac{2 [ 1 - \cos ( \vartheta )- \sqrt{3} g \sin ( \vartheta) ]  }{ 3 [ 1 + g^2 + ( g^2 - 1 ) \cos ( 
				\vartheta) ] }
			\:\:\:\: . 
			\label{para.4}
		\end{eqnarray}
		\noindent
		Plugging Eqs.\eqref{para.3} and \eqref{para.4} into  Eqs.\eqref{junct.12G} and \eqref{junct.12K}, we eventually 
		obtain the CCT, given by  \cite{Oshikawa2006}
		\beq
		\mathbb{G} = \frac{e^2 \left[ \left({\bf 1}-3\mathbb{I}\right)t_\vartheta^2
			+\sqrt{3}g\hat\epsilon t_\vartheta\right]}
		{3\pi\left(g^2+t_\vartheta^2\right)}
		\;\;\; , 
		\label{cct.1}
		\eneq
		\noindent
		and the HCT given by  
		\beq
		\mathbb{K} = - \frac{2\pi k_B^2 T t_\vartheta^2}{27\left(g^2+t_\vartheta^2\right)^2}\left[\left(3g^2-
		t_\vartheta^2\right)\left({\bf 1} -3\mathbb{I}\right)+2g\sqrt{3}\hat\epsilon t_\vartheta\right]
		\:\:\; .  
		\label{para.6}
		\eneq
		\noindent
		In Eqs. \eqref{cct.1} and \eqref{para.6}, the various tensors are
		defined so that ${\bf 1}_{i , j } = 1$ , $\mathbb{I}_{i,j} = \delta_{i,j}$, and $\hat{\epsilon}_{j,k}=\sum_l\epsilon_{jkl}$.
		Also, we have set  $t_\vartheta=\tan\frac{\vartheta}{2}$. 
		Taking the ratio between the entries of the conductance tensors with the same pair of indices, 
		we obtain the Lorenz ratio 
		\begin{eqnarray}
			\mathcal{L}_{j,j'} &=& 2\frac{3g^2 \delta_{j,j'}+t_\vartheta^2+\sqrt{3}g t_\vartheta\epsilon_{j,j'}}{3\left(g^2+t_\vartheta^2\right)}L_0\:\: . 
			\label{para.7}
		\end{eqnarray}
		\noindent
		Our derivation of Eqs. \eqref{cct.1} and \eqref{para.6} relies on the existence of the scale invariant matrix ${\bf \rho}$, characterizing a FP in the phase diagram of the system.  Nevertheless, due to the symmetry of the $\rho$ matrix, a generalization of  Eqs. \eqref{cct.1} and \eqref{para.6} is expected to hold even outside of the FPs, provided that Kirchhoff law for the thermal and for the charge currents is valid and that the boundary interaction Hamiltonian is symmetric under swapping any two leads with each other and exchanging $\vartheta$ with $2 \pi - \vartheta$. In this case, we expect  $G_{j,j'}$ and $K_{j,j'}$ to take the general expression 
		${\cal A} ( g , \vartheta , D ) (1 - 3 \delta_{j,j'} ) + {\cal B}( g , \vartheta , D ) \hat{\epsilon}_{j,j'}$, with 
		${\cal A} , {\cal B}$ being functions of $g$, $\vartheta$ and of a running dimensionful   energy scale $D$ (which in 
		the following we identify with $k_B T$).  
		
		Admitting explicit breaking of $\mathbb{Z}_3$-symmetry, while still requiring charge conservation
		allows another class of splitting matrices \cite{Bellazzini2007,Bellazzini2009}, distinct from the one in Eq. \eqref{para.3}, \eqref{para.4}. 
		In this case, ${\bf \rho}_{B}$ does not depend on $g$,   as it can be readily checked using the 
		formalism of Appendix \ref{Renormalization}, and its general form is 
		\beq
		{\bf \rho}_B  =
		\left( \begin{array}{ccc}
			\hat{b} ( \vartheta ) & \hat{a} ( \vartheta )     & \hat{c} ( \vartheta )   \\
			\hat{a} ( \vartheta ) & \hat{c} ( \vartheta )     & \hat{b} ( \vartheta )   \\
			\hat{c} ( \vartheta ) & \hat{b} ( \vartheta )     & \hat{a} ( \vartheta ) 
		\end{array} \right)
		\;\;\;\; , 
		\label{para.9}
		\eneq
		\noindent
		with $\hat{a} ( \vartheta ), \hat{b} ( \vartheta ), \hat{c} ( \vartheta ) $
		obtained from $ a ( \vartheta ) , b ( \vartheta ) , c ( \vartheta )$, respectively, in 
		Eq.\eqref{para.4} by setting $g=1$
		\footnote{Our parametrization differs by the one in \cite{Bellazzini2009} by a redefinition $\theta\to\pi/3-\theta$}.
		
		The  conductance tensors are directly obtained from 
		Eqs.\eqref{para.6} and \eqref{para.7} by means of the replacement \mbox{$j'\to 3- j'$}.
		Accordingly, we  now obtain for the Lorenz ratio
		\beq
		\mathcal{L}_{j,j'} = \frac{1}{3}-\delta_{j,j'}+\frac{2}{3}\cos\left[\vartheta+\frac{2\pi(j + j')}{3}\right]
		\:\:\:\: . 
		\label{para.10}
		\eneq
		\noindent
		An alternative situation of physical interest is the one in which a ``dual'' Kirchhoff law holds, in that the {\it total charge} entering/exiting the junction is equal to zero \cite{Bellazzini2009a}.
		Physically, this corresponds to having only Andreev-like scattering processes at the central island (regardless of whether they are single-, or multi-particle), that is, any incoming charge from a lead exits toward either the same or any other lead, as the same charge with opposite sign. 
		By swapping the current and the charge operators with each other 
		(this is equivalent to changing the sign of the chiral $\varphi_{L,j}$ fields, while leaving the one of 
		the $\varphi_{R , j}$ fields unchanged). 
		one obtains the corresponding splitting matrices \cite{Das2008}
		\beq
		{\bf \rho}_A ( \vartheta ) = - {\bf \rho} ( \vartheta )\;\; ,
		\label{para.11}
		\eneq
		with $ {\bf \rho}$ in Eq.\eqref{para.3}, with a similar relation holding for 
		$\rho_B$. The charge conductance takes the form
		\beq
		G_{j,j'}= -\frac{e^2}{\pi}\frac{3g^2 \delta_{j,j'}+t_\vartheta^2+\sqrt{3}g
			t_\vartheta\epsilon_{j,j'}}{3\left(g^2+t_\vartheta^2\right)}\;\;, \label{paraq.1}
		\eneq
		\noindent
		while the heat conductance is still given by Eq.\eqref{para.7}. Finally, the Lorenz ratio is given by
		\beq
		\mathcal{L}_{j,j'}=
		-\frac{2\left[ \left(1-3 \delta_{j , j'} \right)t_\vartheta^2
			+\sqrt{3}g \epsilon_{j,j'}  t_\vartheta\right]}
		{3\pi\left(g^2+t_\vartheta^2\right)}L_0\;.\label{paraq.2}
		\eneq
		\noindent
		For $\vartheta = 0$, Eqs. \eqref{paraq.1} and \eqref{paraq.2} describe the $D^3$ FP (see Section \ref{dpn} below).
		
		\section{Fixed points of junctions of $N$ interacting quantum wires}
		\label{sec:Examples}
		
		We now apply the formulas of Section \ref{sec:Conductance_tensor} 
		to  compute the CCT and the HCT at several fixed 
		points of a junction of QWs, characterized by  conformal boundary conditions  such as the 
		ones in Eq.\eqref{eq:rho}. We work with a junction with a generic number $N>2$ of leads 
		\cite{Nayak1997,Bellazzini2007,Bellazzini2008,Bellazzini2009}, whereas we eventually address specific examples with $N=3$ 
		\cite{Oshikawa2006,Hou2012,Meyer2021,Aristov2011,Rahmani2010,Aristov2017,Das2008,Tokuno2008}.
		Assuming over-all charge conservation at the junction, generalizing the construction of   \cite{Oshikawa2006} 
		to the $N$ QW junction, we introduce the
		center-of-mass $\Phi$ and the relative fields $\xi_a ( x )$ ($a=1 , \ldots , N-1$) \cite{Eriksson2014}
		\beq
		\left( \begin{array}{c}
			\Phi ( x ) \\ \xi_1 ( x ) \\ \vdots \\ \xi_{N-1} ( x )         
		\end{array} \right) = {\bf M}_N \cdot 
		\left( \begin{array}{c}
			\phi_1 ( x ) \\ \phi_2 ( x ) \\ \vdots \\ \phi_N ( x )         
		\end{array} \right) 
		\:\:\:\: , 
		\label{ren.8}
		\eneq
		\noindent
		with the orthogonal matrix
		\beq
		{\bf M}_N = 
		\left( \begin{array}{cccc}
			\frac{1}{\sqrt{N}} & \frac{1}{\sqrt{N}} & \ldots & \frac{1}{\sqrt{N}} \\
			\frac{1}{\sqrt{2}} & - \frac{1}{\sqrt{2}} & \ldots & 0 \\
			\ldots & \ldots & \ldots & \ldots \\
			\frac{1}{\sqrt{N ( N-1 ) }} & \frac{1}{\sqrt{N ( N-1 ) }} & \ldots & - \frac{N-1}{\sqrt{N ( N-1 ) }}
		\end{array} \right) 
		\:\:\:\: , 
		\label{ren.9}
		\eneq
		\noindent
		(and similar ones for the $\theta_j$ fields). Charge conservation at the junction implies that $\partial_x\Phi ( 0 )=0$  \cite{Oshikawa2006}. At the 
		disconnected fixed point (DFP), at which all the QWs are disconnected from each other, 
		also the $\{ \xi_a ( x ) \}_a$ obey Neumann boundary conditions, which is equivalent to Eqs.\eqref{exdis.1} below.
		
		A simple way for constructing ``nontrivial'' FPs with alternative conformal boundary conditions is to trade the boundary conditions in one or more combinations of fields 
		from Neumann to Dirichlet. Pertinently imposing  Dirichlet boundary conditions in the relative channels, it is possible to construct FPs
		characterized by multiparticle scattering processes at the central island. Following the discussion of   Section \ref{sec:Noninteracting},
		we expect these FPs to be good 
		candidates to host a violation of the WFL.

		\subsection{The disconnected junction}
		\label{subsec:Disconnected}

		The DFP describes disconnected QWs, which  is
		accounted for by imposing open boundary conditions on the system
		\beq
		\rho_{j,j'} = \delta_{j , j' } 
		\:\:\:\: ,
		\label{exdis.1}
		\eneq
		\noindent
		which corresponds to setting $\vartheta = 0$ in the right-hand side 
		of Eq.\eqref{para.4}.
		Accordingly, $G_{j,j'} = K_{j,j'} = 0$ for all pairs of indices.
		While the result at the fixed point is in itself trivial, we  employ the 
		conditions in Eq.\eqref{exdis.1} to write boundary perturbations to the DFP in Section \ref{sec_Renormalization_3}.

		\subsection{The chiral fixed points}
		\label{chfp}
		
		In the noninteracting, $g=1$ limit, we characterize   the  chiral FPs  $\chi^\pm$ by the boundary conditions 
		\beq
		\varphi_{R , j} ( 0 ) = \varphi_{L , j \pm 1} ( 0 ) 
		\;\;\; , \;\; (j+N \equiv j) 
		\;\;\;\; . 
		\label{chfp.1}
		\eneq
		\noindent
		The corresponding splitting matrix is $ \rho_{j,j'}=\delta_{j,j'\pm1}$. 
		Physically, this corresponds to perfect transmission of a particle entering 
		from lead $j$ into lead $j \pm 1$, and to zero transmission amplitude into any 
		other lead.  For $g \neq 1$, the splitting matrix is constructed 
		using the formulas of Appendix \ref{Renormalization}. For 
		$N=3$, one obtains \cite{Oshikawa2006}
		\beq
		{\bf \rho}_{\chi_\pm} = \frac{2 }{3 + g^2}\left( \begin{array}{ccc}
			- \frac{1 - g^2}{2} &  1 \pm g &     1\mp g    \\
			1\mp g   & - \frac{1 - g^2}{2}  & 1 \pm g \\ 
			1 \pm g &    1\mp g  & - \frac{1 - g^2}{2} 
		\end{array} \right)
		\;\;\;\; . \label{junct.14}
		\eneq
		\noindent
		Eqs.\eqref{junct.12G} and \eqref{junct.12K} then yield
		\begin{eqnarray}
			{G}^{\chi_\pm }_{j,j'}  &=&\frac{e^2 g}{\pi\left(3+g^2\right)}\left[1-3\delta_{j,j'} \pm g\epsilon_{j,j'}\right]
			\:\:\:\: ,
			\label{chfp.5}  \\
			K^{\chi_\pm}_{j,j'} & =& \frac{2\pi k_B^2 T}{3\left(3+g^2\right)^2} \left[\left(1+g^2\right)\left(1-3\delta_{j,j'}\right)\pm 2g\epsilon_{j,j'}\right] . 
			\label{chfp.6}
		\end{eqnarray} 
		\noindent
		By inspection, the results in Eqs.\eqref{chfp.5} and \eqref{chfp.6} imply a violation of the WFL for $g\ne1$, encoded in the 
		(renormalized) Lorenz ratio
		\beq
		\mathcal{L}_{j,j'}=2\frac{1 + g^2\delta_{j,j'}\pm g \epsilon_{j,j'}}{3+g^2}L_0
		\eneq
		Nevertheless the violation is only due to the interaction in the leads and, as discussed in 
		Appendix \ref{Renormalization}, it disappears when the junction is connected to Fermi liquid  reservoirs. 
		In this case $\hat{\bf \rho}_{\chi_\pm} = ({\bf 1} \pm \hat\epsilon-\mathbb{I})/2$ and the conductance 
		tensors are directly obtained from Eqs.\eqref{junct.14} and \eqref{chfp.5} by setting $g=1$.  As stated above, our conclusion applies  only provided that the electric conductance tensor component is different from zero. 
		The above discussion clearly applies to the $\chi^\pm$ FPs of a generic junction, with any number of leads $N$ and straightforwardly generalizes to any splitting matrix which represents a permutation, 
		i.e., a matrix which has all vanishing entries, except for 
		one off-diagonal entry in each row and in each column. This is the only situation in which the boundary 
		conditions can be  equivalently formulated in fermionic variables, \mbox{$a_{j',j} = r_{j',j} = c_{j',j} = 0$},
		\mbox{$\forall j',j= 1 , \ldots , N$}, and $t_{j',j}= \delta_{j'\pm1,j}$ for the $\chi^\pm$  FPs. 
		Apparently, in this case the dynamics is described by single-particle scattering processes only.
		
		For $N=3$, the $\chi^\pm$ FPs are stable for $1<g<3$ \cite{Oshikawa2006}. At variance, for $N \geq 4$ the $\chi^\pm$ 
		FPs are unstable for any value of $g$.   
		
		\subsection{The $D^{N-1}$ fixed point}
		\label{dpnfp}
		
		We consider the FP recovered by imposing Dirichlet boundary conditions onto all the relative fields in Eq.\eqref{ren.8}, which we generically 
		dub $D^{N-1}$. In the $N=3$ junction, two different types of FPs emerge, denoted by $D_P$ and $D_N$ in \cite{Oshikawa2006}: they share the same splitting matrix, but differ by their operator content and have therefore a different range of stability in the system parameters. 
		Accordingly, while they have the same FP conductance tensors, the finite-temperature corrections will be different, as they scale with a power law of the temperature, which depends on the dimension of the leading irrelevant operator.
		
		The corresponding FP splitting matrix is given by \cite{Eriksson2014_PRL,Eriksson2014}
		\beq
		{\bf \rho}_{D^{N-1}} = \frac{2}{N} {\bf 1}  -  \mathbb{I}
		\:\:. 
		\label{ren.10}
		\eneq
		\noindent
		Computing the charge and heat conductance tensors with the formalism of Section \ref{sec:Conductance_tensor}, we obtain 
		\beq
		\mathbb{G} = \frac{e^2 g}{\pi} \: \left(   \frac{1}{N}  {\bf 1}  -  \mathbb{I} \right)\quad,\quad
		\mathbb{K} = \frac{2 \pi k_B^2 T}{3N} \left( \frac{1}{N}  {\bf 1} -  \mathbb{I}  \right)
		\:\:. \label{ren.11}
		\eneq
		\noindent  
		Eq.\eqref{ren.11} is obtained from Eqs.\eqref{cct.1} and \eqref{para.6} by setting $\vartheta = \pi$. 
		From   Eq.\eqref{ren.11} we directly read  
		\beq
		\mathcal{L}_{j,j'} =  \frac{2}{N} \frac{\pi^2  k_B^2  }{3 g e^2}
		\:\:\:\:.
		\label{ren.12}
		\eneq
		\noindent 
		In Eq.\eqref{ren.12} we identify with the factor $g^{-1}$ the contribution merely stemming from the bulk interaction in the QWs
		(which is washed out  once the junction is connected to Fermi liquid reservoirs, see Appendix \ref{Renormalization} for details) and the 
		factor $\frac{2}{N}$ due to the multi-particle scattering processes at the central island. 
		As the renormalization of the Lorenz ratio is present even when the junction is connected to external  Fermi liquid  leads,
		we conclude that the WFL breaks down at the $D^{N-1}$ FP, which highlights that the dynamics  cannot be described within the 
		single-particle framework of Section \ref{sec:Noninteracting}. 
		In fact, it is directly related to the onset of zero-temperature multi-particle scattering processes at the central island, consistently 
		with the results of \cite{Nayak1997}, where a similar phenomenon was studied at a resonant, multi-lead quantum point contact. 
		Given the ${\bf \rho}$-matrix in Eq.\eqref{ren.10}, we may readily identify a scattering process in which $N$ particles are injected 
		into the central island from, e.g., lead $j$. The incoming particles are symmetrically transmitted into the remaining  leads as $N-1$ pairs, each of charge $2e$.
		As charge is conserved, $N-2$ holes are Andreev backscattered into lead $j$. 
		Alternatively, we may consider a ``dual'' process in which, e.g., two particles (total charge $2e$) are  injected from each lead $j'$, with $j' \neq j$, and a mix of normal transmission and crossed Andreev reflection yields $N$ particles and $N-2$ holes exiting the central island from lead $j$. In
		Fig.\ref{multi_scattering} we draw a sketch of the two processes. 
		\begin{figure}
			\center
			\includegraphics*[width=.8 \columnwidth]{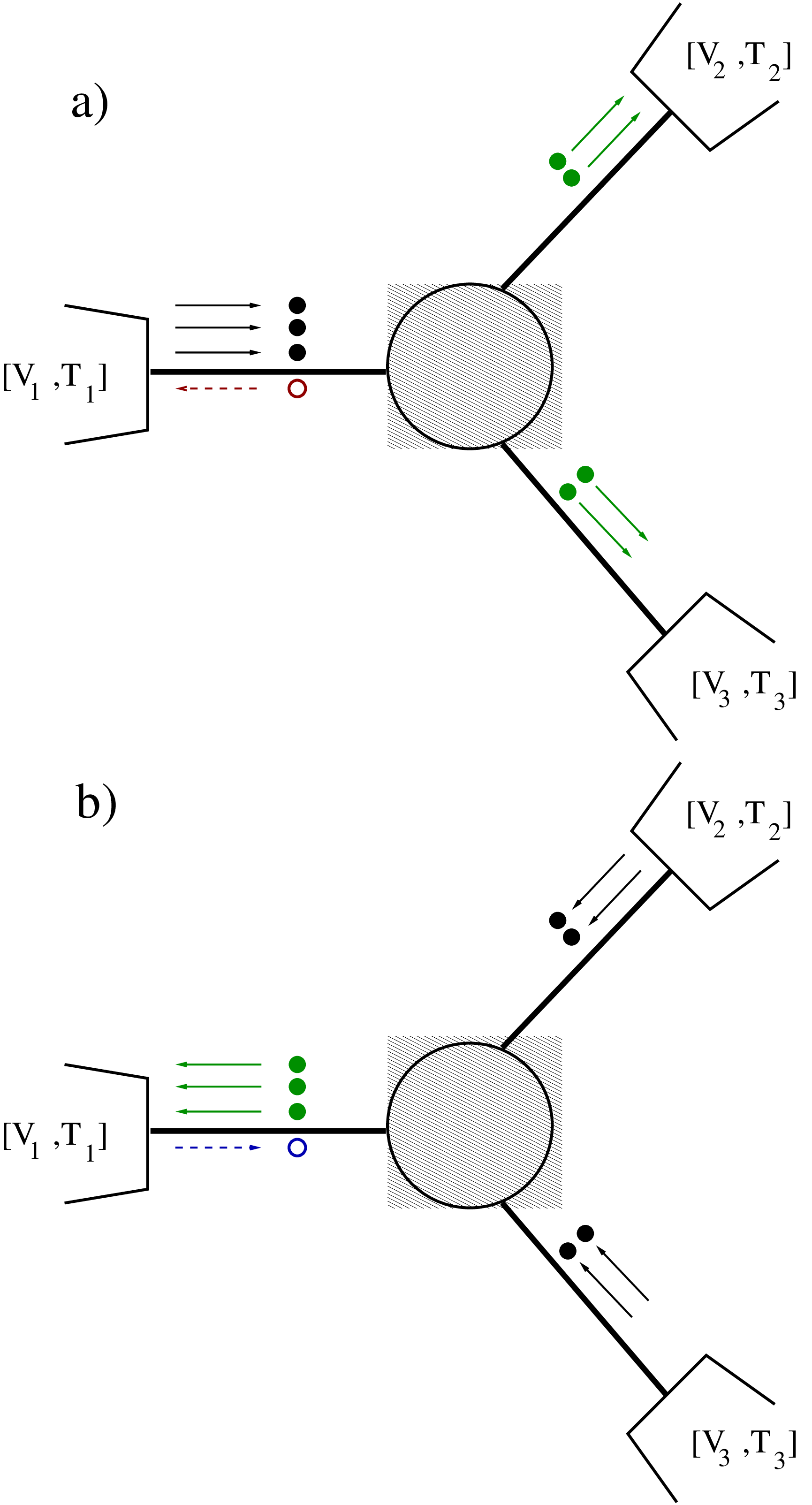}
			\caption{Sketch of two multiparticle scattering processes taking place at 
				the $D^2$ FP of the $N=3$ junction  (for $g=1$). Specifically, \\
				{\bf a)}: three particles (black full dots) are injected into the central island from lead 1. Two pairs of particles are  symmetrically transmitted
				into leads 2 and 3 (green full dots) while, consistently with the total charge conservation,  a hole is Andreev backscattered into 
				lead $1$ (red open dot). \\
				{\bf b)}:  Two particles are  injected from leads 2 and 3 (black full dots) 
				and a mix of normal transmission and crossed Andreev reflection yields an outgoing, multiparticle state within lead 1, 
				consisting of $3$ particles (green full dots)  and one  hole (blue open dot).  }
			\label{multi_scattering}
		\end{figure}
		\noindent
		Alternatively, we may borrow the second point of view of 
		\cite{Nayak1997} by considering a single-particle  
		``in'' state that, consistently with the picture of Fig.\ref{multi_scattering}, 
		gives rise to ``out'' states in the other leads whose charge is, in 
		general, no longer a multiple of the unit charge ($2e/N$ in our specific 
		model). This charge fractionalization under scattering at the central island marks an apparent breakdown of the Fermi liquid 
		picture and works as a pictorial  explanation of the breakdown of the WFL.
		
		We expect that a pertinent renormalization of the Lorenz ratio also takes place, for $N \geq 4$, at reduced symmetry FPs, such as the $D^{N-2}$FP, 
		where Dirichlet boundary conditions are imposed to only $N-2$ of the relative fields, while the remaining one 
		obeying Neumann boundary conditions, together with the center-of-mass field.

		\subsection{The $D^N$ fixed point}
		\label{dpn}
		
		Compared to the $D^{N-1}$ FP, the $D^N$ FP is characterized by the center-of-mass  field  $\Phi$ defined in Eq.\eqref{ren.8} satisfying 
		Dirichlet, instead of Neumann,  boundary conditions. Correspondingly, the  charge conservation breaks down and the WFL can be violated, even though the corresponding FP can be fully described in terms of single-particle scattering processes only.
		The $D^N$ FP is dual (in the sense of the charge-current duality \cite{Das2008}) of the 
		DFP, thus, the splitting matrix is given by ${\bf \rho} = -\mathbb{I}$ and describes perfect Andreev reflection in each lead \cite{Eriksson2014_PRL,Eriksson2014}.
		In the language of Section \ref{sec:Noninteracting},  $A_{j,j} = 1$ $\forall j$, all the other scattering coefficients being zero. This implies $G_{j,j'} = \frac{e^2}{\pi} \delta_{j,j'}$ and $K_{j,j'} = 0$, $\forall j,j' = 1, \ldots , N$.
		In fact, this is just what happens, at low enough energies, in a single wire coupled to a topological superconductor \cite{Fidkowski2012,Affleck2013,Affleck2014}.
		As a result, Andreev reflection becomes a resonant process at the FP, with no room left 
		for normal, single-particle backscattering. 
		In general, when considering the heat conductance through a normal metal-superconductor interface, we  find that it strongly depends on the mechanism of electron transfer across the interface.  Specifically, when single-particle transfer dominates and, accordingly, the backscattering on the normal side of the interface is mostly normal, electronic states on the normal side at energies above the superconducting gap are depleted, resulting in an effective cooling mechanism of the metal. At variance, when the coherent two-electron tunneling becomes the dominant mechanism 
		for charge transfer across the junction (corresponding to the onset of Andreev reflection on the normal side), the heat flow is strongly suppressed, due to the fact that now electrons with all energies, including those inside the energy gap, are removed from the normal
		metal \cite{Averin1995}. Accordingly, we expect that the full suppression of normal backscattering versus Andreev reflection implies a full suppression of the heat flow through the interface and, in addition, that this conclusion holds regardless of whether the superconducting side is topological.
		
		In conclusion, we may regard the $D^N$FP as "trivially" violating the WFL. The violation is, indeed, just related to the peculiar  subgap physics of the NS-interface. For this reason, in the following we focus  on  the charge conserving junction with $N=3$.
		
		\section{Phase diagram and transport in charge-conserving $N=3$ junctions}
		\label{sec_Renormalization_3}
		
		As a specific example  of realization  of the fixed points described above, we discuss in detail the $N=3$ junction of interacting QWs. Specifically, in 
		the following  we focus on two types of boundary interactions: the direct fermion hopping between 
		lead ends in  \cite{Chamon2003,Oshikawa2006} and the TKM 
		discussed in \cite{Beri2012,Altland2013}.
		
		The simplest, nontrivial example is the $N=3$ junction discussed in Refs.\cite{Chamon2003,Oshikawa2006}, whose  generalization to a generic 
		$N (\geq 4)$ is presented in \cite{Bellazzini2008}. In such a system, once resorting to the bosonization framework, in terms of the 
		unfolded chiral fields defined in Eq.\eqref{6}, the boundary interaction at the DFP is given by 
		\beq
		H_{{\rm Junc}, N} = - J_K \sum_{k<l =  1}^N \Gamma_k \Gamma_l \: e^{ i \sqrt{\frac{4 \pi}{g}} 
			[ \varphi_k ( 0 ) - \varphi_l ( 0 ) ] + i \chi_{k,l} } + {\rm h.c.} 
		\:\:\:\: , 
		\label{exdis.2}
		\eneq
		\noindent
		with $J_K$ being  the over-all boundary coupling strength and the $\chi_{k , l}$ being phases that may enter $H_{{\rm Junc} , N}$ if, e.g., 
		there is a magnetic flux piercing the junction itself \cite{Oshikawa2006}. The Hamiltonian in Eq.\eqref{exdis.2} conserves the total charge, 
		but breaks time reversal invariance  for a generic choice of the phases $\chi_{k,l}$. 
		In general, it is a relevant boundary operator as soon as  $g>1$.
		
		In the TKM, a superconducting island is present at the junction, hosting low-energy degrees of freedom in the form of MZMs, which are in turn 
		tunnel-coupled to the end of the leads. The superconducting island itself is floating and characterized by a large charging energy $E_c$, 
		which ultimately determines 
		the charge conservation at the junction. The boundary Hamiltonian describing such a system in the cotunneling regime is
		\beq
		H_{{\rm TK} , 2} =  - 2   \sum_{ k < l = 1}^N J_{k,l}  
		\cos \left[ \sqrt{\frac{4 \pi}{g}}  (  \varphi_k ( 0 ) -  \varphi_l ( 0 )) +\chi_{k,l}\right]
		\:\:\:\: ,
		\label{eq.10}
		\eneq
		with $J_{k,l} \sim  1 / E_c$. No Majorana fields, nor Klein factors appear in Eq.\eqref{eq.10}, due to the ``Majorana-Klein hybridization''  \cite{Altland2013,Beri2013,giuliano2019}, 
		which factors them out of the dynamics. The boundary term can be regarded  as a generalization of the Kondo model to the $SO(N)$ symmetry group \cite{Beri2012,Altland2014,Buccheri2015}. 
		$H_{{\rm TK} , 2}$ is relevant for $g>1$ and marginally relevant for $g=1$ \cite{Beri2012,Altland2013,Cardy1996,giuliano2018,giuliano2019,Kane2020}. Anisotropy in 
		the $J_{k , l}$ are washed out along the renormalization group (RG) trajectories. Accordingly, without  any loss of generality, 
		from now on we assume  $J_{k , l}=J_K$ for every pair of wires.
		
		In a related setting, the superconducting island can be Josephson-coupled to another superconductor, which breaks charge conservation at the junction. In this case, 
		the boundary Hamiltonian is given by \cite{Eriksson2014_PRL,Eriksson2014}
		\beq
		H_{{\rm TK} , 1} = -  \sum_{ j = 1}^N  \sqrt{2} t_j  \sin \left[  \sqrt{\frac{4 \pi}{g}}  \varphi_j ( 0 ) \right] + H_{{\rm TK} , 2}
		\label{eq.8} 
		\:\: ,
		\eneq
		with $t_j\sim E_J$ where $E_J$ denotes the Josephson energy. The first term always triggers a flow toward a FP at which $\phi_j ( 0 )$ is pinned to some 
		nonuniversal value, depending on the specific ``bare'' values of the boundary interaction strengths \cite{Eriksson2014_PRL,Eriksson2014}. In the ``phase'' 
		regime $E_c \ll E_J $, the low-temperature FP is known as $SO(N)_1$ Topological Kondo  FP. The first term in Eq.\eqref{eq.8} has  
		scaling dimension $(2g)^{-1}$ and is therefore  relevant as soon as $g> \frac{1}{2}$.
		
		We now investigate in detail the various phases with the corresponding transport properties. 
		
		\subsection{The disconnected fixed point}
		\label{discfip_3}
		
		As we discuss in Section \ref{sec:Examples}, at the DFP one finds vanishing conductance tensors. Turning on the boundary interaction, we perturbatively compute 
		the expectation values of the currents by employing the Keldysh approach of Appendix \ref{General:expression}. As a result, we obtain for the 
		electric and thermal conductance tensors \cite{Us_short2021}
		\begin{eqnarray} 
			G_{j , j'} &=&   
			\frac{6 \pi e^2 \Gamma^2 \left( \frac{1}{g}  \right) {\cal J}^2\left(  D \right) }{  \Gamma   \left( 2 /g \right) }
			\left( \frac{1}{3} -   \delta_{j , j'}  \right) , \label{Gpert} \\
			K_{j , j'} &=& \frac{2 \pi^3 k_B^2 T \Phi(g)  {\cal J}^2\left( D \right)   \Gamma^2 \left( \frac{1}{g} \right) }{  \Gamma   \left( 2 /g \right)   }
			\left( \frac{1}{3} -   \delta_{j , j'}  \right),
			\label{Kpert} 
		\end{eqnarray}  
		\noindent
		with the dimensionless running coupling  \mbox{${\cal J} (D) =  \frac{ J_K}{D_0}  \left( \frac{D}{D_0} \right)^{-1 + \frac{1}{g}}$},
		$D=2 \pi k_B T$  a scale with the dimension of an energy,
		and $D_0$ a high energy cutoff.
		The latter is a relevant large energy scale, such as the bandwidth of the conduction band or the charging energy of the floating island in the TKM example in Sec. \ref{phadiag} below. 
		As discussed above, due to the system symmetries, we expect that  $G_{j,j'} = {\cal A}_G ( g , \vartheta , D ) (1 - 3 \delta_{j,j'} ) + 
		{\cal B}_G( g , \vartheta , D ) \hat{\epsilon}_{j,j'}$  and $K_{j,j'}= {\cal A}_K ( g , \vartheta , D ) (1 - 3 \delta_{j,j'} ) + 
		{\cal B}_K ( g , \vartheta , D ) \hat{\epsilon}_{j,j'}$. As it appears from the right-hand side of Eqs. \eqref{Gpert} and \eqref{Kpert}, this is indeed the case.
		As a consequence,  the two conductances have the same tensor structure, hence, the ratio between any pair of  nonzero  entries is
		\beq
		\mathcal{L}_{j , j'} = \Phi ( g ) L_0
		\:\: , 
		\label{n3.3}
		\eneq
		with
		\begin{eqnarray} \label{n3.2}
			&& \Phi ( g ) =\frac{3  \Gamma \left( 2/g \right)}{g \pi \Gamma^4 \left( 1/g \right)}
			\int d z d w  \frac{z}{\sinh ( \pi z ) }  
			\\
			&&  \qquad\qquad \times \left| \Gamma \left( \frac{1}{2 g} +  i \left( z    - w \right) \right) \Gamma \left( \frac{1}{2g} +  i  w \right) \right|^2  
			\:\: . \nonumber 
		\end{eqnarray}
		\noindent
		As expected, $\Phi ( g = 1 ) = 1$, which can be shown using the identity \eqref{dfp.2.13} in  Appendix \ref{General:expression}. When $g \neq 1$ but $| g - 1 | \ll 1$, we may improve the results 
		in Eq.\eqref{n3.2} by letting    ${\cal J} (D)$
		flow with the running energy scale $k_B T$ according to the appropriate RG equations (see Appendix \ref{app:RGTKM} for details).
		Within the perturbative approach to the TKM, the main effect is the scaling of 
		both $G_{j,k}$ and $K_{j,k}$ with the running coupling. Accordingly, the Lorenz ratio is scale-independent and equal to $L_0$ for $g=1$.
		An important difference between the TKM \eqref{eq.10} and the $N=3$ junction \eqref{exdis.2} emerges for $g=1$, as the boundary interaction in the $N=3$ junction is purely marginal \cite{Oshikawa2006}. In this case, Eqs. \eqref{Gpert},\eqref{Kpert} and \eqref{n3.2} provides the leading perturbative (in $J_K$) contributions to the conductance tensors at a manifold of Fermi liquid fixed point, consistently with the result $\Phi (1) = 1$.
		Conversely,  the boundary interaction \eqref{eq.10} is marginally relevant for $g=1$: in this case, we cannot rely on the above results at low temperatures, but we rather need to assess the stable fixed point encoding the $T \to 0$ behavior of the junction.  
		To do so, we now go through an extensive  review of the phase diagram of the $N=3$ junction and of the TKM.

		\subsection{Phase diagram of the $N=3$ junction and of the  topological Kondo model}
		\label{phadiag}
		
		While our derivation allows us to make a sharp prediction on the value of the Lorenz ratio, in order for the effect to be robust in a realistic system, the $D^{N-1}$FP has 
		to be an infrared attractive RG fixed point. In junctions of normal QWs, this happens only at rather unphysically large values of the (attractive)
		interaction strength in the leads \cite{Chamon2003,Oshikawa2006}.
		The $D^{N-1}$ FP also emerges in the phase diagram of the TKM  \cite{Beri2012,Altland2013,Eriksson2014}: remarkably, it is stable as soon as $g>\frac{N}{2(N-1)}$ \cite{Beri2012,Altland2013}. 
		This points toward an intriguing relation between the emergence of MZMs and the detection of a robust violation of the WFL as discussed above, so that the latter 
		effect may be used as an evidence for the presence of MZMs. In order to better spell out this point, we discuss the case $N=3$, for which a complete classification of the  FPs and of the corresponding conformal boundary conditions is possible \cite{Oshikawa2006,Affleck2001}.
		We assume for simplicity $\chi_{l,l+1}=\chi/3$, see Eqs. \eqref{exdis.2} and \eqref{eq.10}, and $-\pi<\chi\le\pi$ throughout this section.
		
		\subsubsection{Direct hopping}
		\label{dhop}
		
		The phase diagram of the $N=3$ junction with direct hopping between the leads Eq.\eqref{exdis.2} has been discussed in detail in \cite{Oshikawa2006}. 
		For $g<1$ (repulsive interaction in the leads), the DFP is infrared (IR) stable. In the absence of interaction, $g=1$, the junction has a manifold of marginally equivalent FPs,
		which can be described in terms of the single-particle ${\bf S}$-matrix 
		approach of Section \ref{sec:Noninteracting}. 
		For $1<g<3$, the system flows instead outside of the weakly coupled regime. Any $\chi \neq 0 , \pm \pi$ breaks time-reversal invariance, triggering a nontrivial renormalization 
		toward either one (depending on the sign of $\chi$) of the chiral FPs of sec. \ref{chfp}, which  are stable as long as $g<3$. For the sake of our discussion, 
		it is useful to remind that, in the noninteracting limit, the chiral FPs $\chi_\pm$  can be described within the single-particle ${\bf S}$-matrix formalism as well. 
		When $\chi=0$, the RG flow points instead toward a time-reversal invariant, finite coupling FP, dubbed ${\bf M}$FP in \cite{Oshikawa2006}. While a full theory of the ${\bf M}$FP is still lacking, based on the numerical results of \cite{Rahmani2010,Rahmani2012}, in the following we argue that the WFL is expected to hold at the ${\bf M}$FP as well, once the junction is connected to external reservoirs. 
		For $g=3$ two disconnected FP manifolds emerge, respectively connected to either one of the chiral FPs, separated from each other by the ${\bf M}$FP. 
		For $3<g<9$ two IR stable strongly coupled FPs emerge in the phase diagram, the $D_P$ FPs, while 
		the $\chi_\pm$FPs disappear. The $D_P$ FPs are still separated by a time-reversal invariant FPs ${\bf M}'$, analogous to the ${\bf M}$FP. Finally, for $g>9$ the ${\bf M}'$FPs disappear, as well.
		Along the time-reversal invariant RG trajectories, the junction flows toward the $D_N$ FP.
		The $D_P$ and the $D_N$ FPs are described by the same ${\bf \rho}$ matrix, that is, 
		${\bf \rho}_{D^2}$ in Eq.\eqref{ren.10} with $N=3$, but differ in their operator content. 
		The $D_P$ FPs are stable for $g>3$, the $D_N$ FPs are stable for $g>9$ \cite{Oshikawa2006}.
		In order to stabilize these FPs, we need a large value of the Luttinger parameter, corresponding to a very strong attractive bulk interaction, hard to realize in realistic junctions.
		
		In Fig.\ref{pd_3}, we draw the RG trajectories of the $N=3$ junction within the various range of parameters discussed above.
		In order to evidence our main conclusions, we highlight as black/green dots the FPs in the phase diagram at which the WFL holds/breaks down, when the junction is  connected to external leads, as we discuss in Section \ref{ct3}.
		
		\begin{figure}
			\includegraphics*[width=1. \columnwidth]{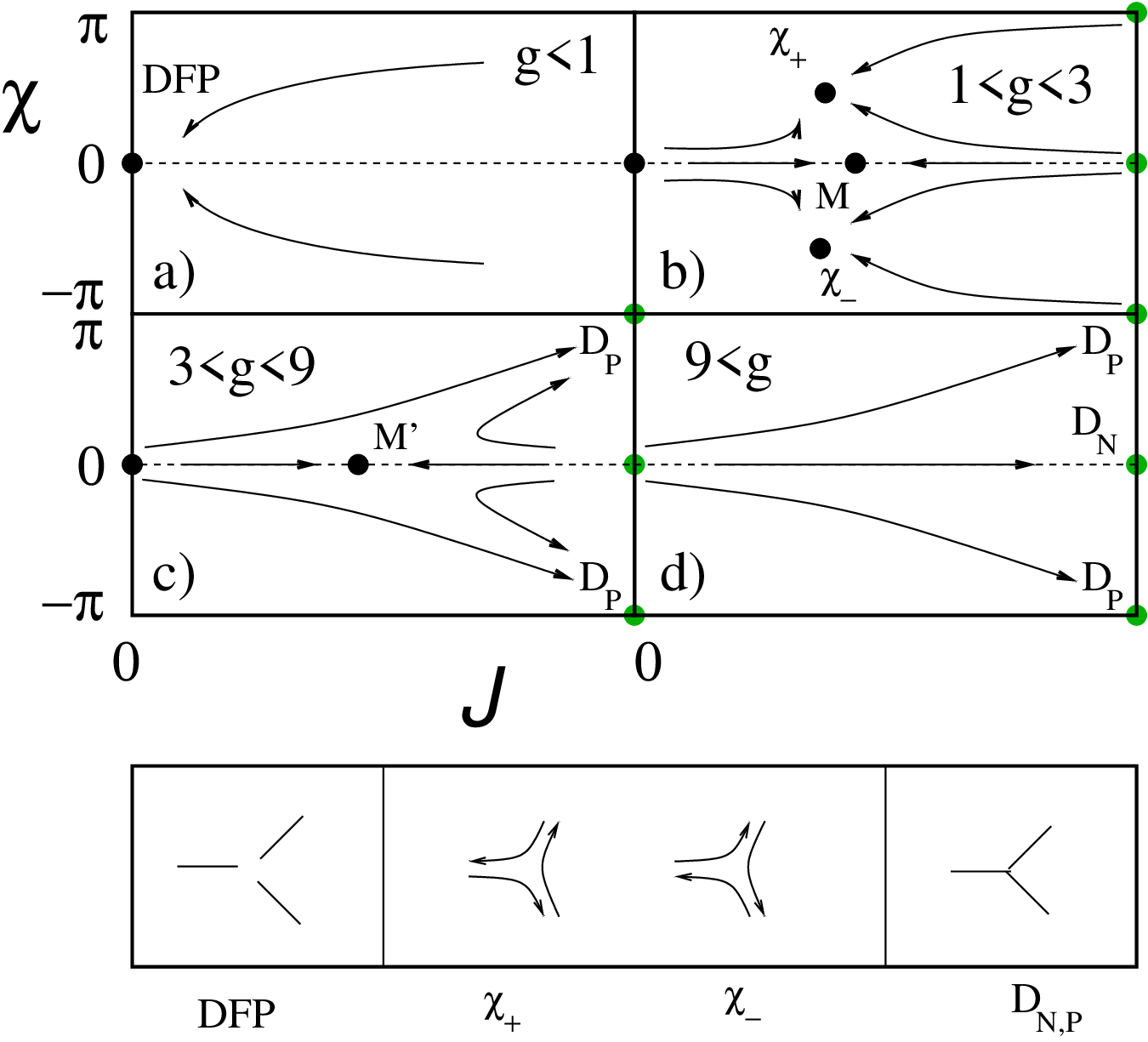}
			\caption{{\bf Top}: Sketch of the RG trajectories of the $N=3$ junction  in the ${\cal J} - \chi$ plane within the relevant ranges of values of 
				the Luttinger parameter $g$. We mark with  green dots the FPs where a violation of the 
				WFL is expected, also when the junction is connected to Fermi-liquid reservoirs.  Specifically, \\
				{\bf a)} $g<1$, only the DFP is stable  and the WFL holds; \\
				{\bf b)} $1<g<3$: depending on the value of $\chi$ at the reference scale, the system flows toward either the ${\bf M}$, or to the $\chi_\pm$ FPs.
				In any of these FPs the WFL holds;  \\
				{\bf c)} $3<g<9$: the $D_P$ FPs become stable and the WFL  breaks down;\\
				{\bf d)} $9<g$: both the $D_P$ and the $D_N$ FPs become stable. In both cases 
				the WFL  breaks down. \\ 
				{\bf Bottom}: Pictorial sketch of the FPs relevant to our analysis. From left to right: the DFP (all the wires are 
				disconnected from each other), the $\chi_\pm$ FPs, and the (fully connected) $D_{N,P}$ FPs.} 
			\label{pd_3}
		\end{figure}
		
		\subsubsection{The Topological Kondo model}
		\label{ttkm}
		
		The RG flow of the TKM has been discussed in \cite{Beri2012,Altland2013}, as well as in \cite{Herviou2016,Michaeli2017}, and the effect
		of a nonzero $\chi$  has been  considered in \cite{Zazunov2017,Giuliano2009}. Here, we summarize the corresponding equations 
		in Appendix \ref{app:RGTKM}.
		For a small "bare" coupling ${\cal J} >0$  at the reference scale $D_0$, the boundary 
		Hamiltonian in Eq.\eqref{eq.10} is relevant for $g>1$. 
		At the same time, for $\chi \neq \pm \pi$, 
		any nonzero $\chi$ renormalizes to zero. For $g \geq 1$, the system flows towards a large-${\cal J}$ FP, which corresponds
		to the charge-conserving $D^2$FP. The corresponding splitting matrix has been determined in \cite{Eriksson2014} and is given in Eq.\eqref{ren.10}
		for $N=3$. With this knowledge, we can define the "unfolded", chiral fields
		\beq
		\tilde{\varphi}_j ( x ) = \Biggl\{ \begin{array}{c}
			\varphi_{L , j} ( x ) \;\;\; , \;\; ( 0 \leq x \leq \ell) \\
			\sum_{j' = 1}^3 [ \rho_{D^2} ]_{j' , j } \varphi_{R , j'} ( - x ) \;\;\; , \;\;
			(-\ell \leq x < 0 ) 
		\end{array}
		\:\:\:\: . 
		\label{rengcom.3}
		\eneq
		In terms of the fields in Eq.\eqref{rengcom.3}, the leading boundary perturbation is given by \cite{Herviou2016}
		\beq
		\tilde{H}_{{\rm TK} , 2} = - 2 h \: \sum_{ j = 1}^3 \: \cos \left[ \frac{4 \sqrt{ \pi g} }{3} ( 2 \tilde{\varphi}_j - 
		\tilde{\varphi}_{j+1} - \tilde{\varphi}_{j-1}) \right] \:\:\:\: , 
		\label{rengcom.4}
		\eneq
		with   $\tilde{\varphi} \equiv \varphi_j ( 0 )$   and $j + 3 \equiv j$. In the formalism of Appendix \ref{General:expression}, Eq.\eqref{rengcom.4}
		corresponds to setting  
		\begin{eqnarray}
			\alpha_{1,2,3}^{2,3} &=& \frac{2 \sqrt{g}}{3} \; ( -2 , 1 , 1 ) \nonumber \\
			\alpha_{1,2,3}^{3,1} &=& \frac{2 \sqrt{g}}{3} \; ( 1, -2  , 1 ) \nonumber \\
			\alpha_{1,2,3}^{1,2} &=& \frac{2 \sqrt{g}}{3} \; ( 1, 1  ,-2  )
			\;\;\;\; . 
			\label{rengcom.5}
		\end{eqnarray}
		\noindent
		The operator in Eq.\eqref{rengcom.4} has scaling dimension $\frac{4g}{3}$. Therefore,  
		the topological Kondo FP for $N=3$ is stable as long as $g> \frac{3}{4}$. Thus, we conclude that the DFP is  attractive 
		as long as $g<1$ and $\chi \neq \pm \pi$, while the topological Kondo FP is  attractive for $g> \frac{3}{4}$ and $\chi \neq \pm \pi$.
		At variance,  for $g \geq 1$ and $\chi \neq \pm \pi$, the system flows toward the topological Kondo  FP described by the splitting matrix ${\bf \rho}_{D^2}$ \eqref{ren.10} and 
		with leading perturbation in Eq.\eqref{rengcom.4}. 
		From the analysis of Appendix \ref{app:RGTKM}, we conclude that $\chi$ does not flow along the fixed lines $\chi = \pm \pi$. 
		In this case, by explicit investigation one finds that 
		the leading boundary interaction at the Topological Kondo FP (which we dub $\hat{D}^2$ in the following) has scaling dimension $\frac{4g}{9}$ \cite{Giuliano2009}.
		In Eq.\eqref{rengcom.6} we provide the explicit formula for the leading boundary perturbation: it has, in fact, 
		scaling dimension $\frac{4g}{9}$ and  is therefore relevant as long as $g \leq \frac{9}{4}$. 
		Accordingly, for $\chi = \pm \pi$, there is a finite window  $1 < g < \frac{9}{4}$  in which both the DFP and the ${D}^2$FP are unstable and there 
		appears a stable, finite coupling fixed point for  RG trajectories originating from both the DFP and the $\hat{D}^2$FP \cite{Giuliano2009}. 
		Analogously to the  {\bf M}FP of \cite{Oshikawa2006}, no complete theory exists for the intermediate-coupling fixed point  and we are so far unable to make a sharp prediction on the corresponding behavior of the HCT and on the possible violation of the WFL.
		
		Also in the regime $g<1$, we recover the RG fixed lines $\chi=0$ and $\chi = \pm \pi$, the former one being attractive, the latter one repulsive. For $\frac{3}{4} < g < 1$, 
		both the weakly coupled DFP and the strongly coupled $\hat{D}^2$FP are stable. This implies 
		the emergence of a repulsive finite-coupling FP in the phase diagram, corresponding to a quantum phase transition between the two of them at ${\cal J}  = \tilde{{\cal J}}_K =\frac{g^{-1}-1}{2}, \chi = 0$  \cite{Herviou2016}.
		
		Finally, we note that, while there is no reason to exclude {\it a priori} the 
		emergence of the $\chi_\pm$FPs in the phase diagram of the TKM, we are not able to 
		recover them as endpoints of RG trajectories fully lying within the ${\cal J}-\chi$ plane,
		differently to what happens in the $N=3$ junction of \cite{Oshikawa2006}.
		Indeed, in order to get access to the time-reversal breaking FPs, one has to introduce an additional, "chiral" boundary interaction, e.g., the analog of the boundary interaction discussed in \cite{Buccheri2018,Buccheri2019} at a $Y$-junction of critical Heisenberg chains, which we do not discuss here.
		
		To summarize the discussion about the TKM, in Fig.\ref{pha_dia_topo}, we draw the RG trajectories for the system for $g< 1$ in Fig.\ref{pha_dia_topo}{\bf (a)} and for $g \geq 1$ in Fig.\ref{pha_dia_topo}{\bf (b)) and {\bf (c)}}.

		\begin{figure}
			\includegraphics*[width=0.9 \columnwidth]{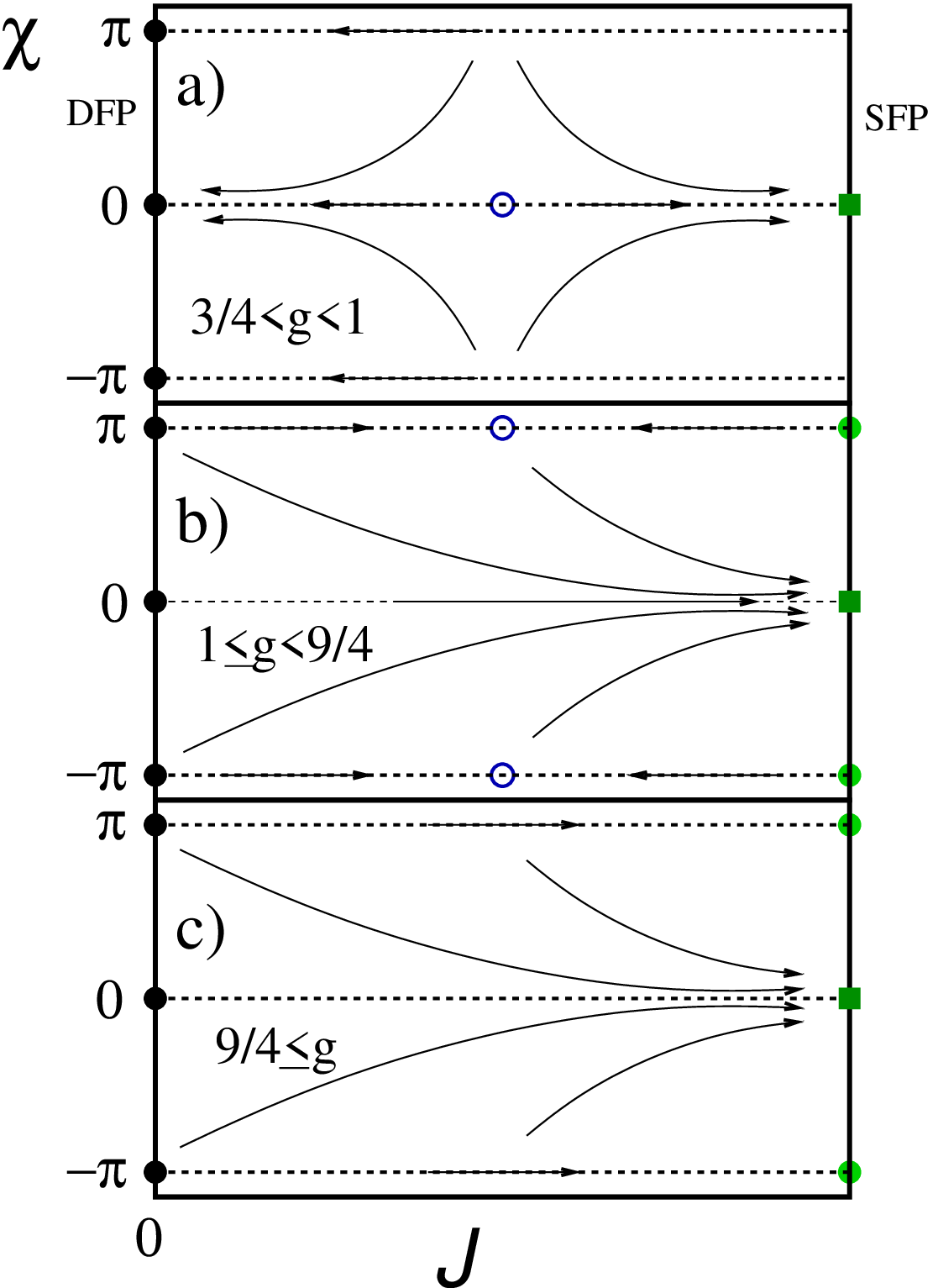}
			\caption{Sketch of the RG trajectories of the TKM in the ${\cal J} - \chi$ plane for various ranges of values of the Luttinger parameter $g$. Here SFP denotes any strong-coupling fixed point.
				The green dots are the FPs where the WFL does not hold for the junction connected to Fermi liquid  reservoirs.
				{\bf (a)} $\frac{3}{4} < g<1$: At $\chi \neq \pm \pi$,  both the  DFP (black dot) and the $D^2$ (dark green square) FPs are stable. 
				The running phase $\chi ( D )$ (see Appendix \ref{app:RGTKM} for details) renormalizes back to the fixed line $\chi = 0$. 
				For certain values of the bulk interactions, a repulsive, finite coupling FP appears (open blue dot).
				\\
				{\bf b()} $ 1 \leq  g < \frac{9}{4}$: For $\chi ( D_0 ) \neq \pm \pi$,   
				$\chi ( D )$ again flows toward the 
				$\chi = 0$ fixed line. ${\cal J} (D)$ is renormalized to strong coupling: the $D^2$FP is stable and, accordingly, the WFL breaks down as $T \to 0$. $\chi = \pm \pi$ 
				are two fixed lines (in $\chi$), along which the DFP and the $\hat{D}^2$ FP are repulsive. The actual stable phase 
				corresponds to the finite coupling FP represented by the open blue dot: its actual nature deserves further investigation;
				\\
				{\bf (c)} $\frac{9}{4} \leq g$: Both the $D^2$FP and the $\hat{D}^2$FP become stable, thus, the WFL breaks down as $T \to 0$, regardless of the initial value of $\chi$.} 
			\label{pha_dia_topo}
		\end{figure}
		\noindent
		
		\subsection{Charge and thermal conductance and the Wiedemann-Franz law in the three-wire junctions}
		\label{ct3}
		
		We now review the charge and the thermal conductance   at the ``nontrivial''
		FPs of the $N=3$ junction and of the TKM  and briefly discuss the scaling properties of the corresponding conductance tensors.
		We begin with  the  {\bf M}FP in the $N=3$-junction: its emergence was originally inferred 
		from the main topology of the phase diagram \cite{Chamon2003,Oshikawa2006}. 
		Later on, it was  confirmed within a combined use of boundary conformal field theory and 
		numerical density matrix RG approach \cite{Rahmani2010,Rahmani2012}, eventually showing that the 
		corresponding CCT for the junction connected to external Fermi liquid  reservoirs is given by 
		\beq
		\mathbb{G}_{\bf M} = \frac{e^2 \gamma}{2 \pi}\left({\bf 1} -3\mathbb{I}\right)\label{ct3.1}
		\:\: , 
		\eneq
		\noindent
		with $\gamma = \frac{4}{9}$, within numerical error bars \cite{Rahmani2012}. 
		Using the RG approach in fermionic coordinates, in 
		\cite{Giuliano2015} it was proposed that the conductance in Eq.\eqref{ct3.1} is determined by 
		the single-particle ${\bf S}$-matrix ${\bf S}_M$, which describes scattering processes at the junction connected to the reservoirs, given by 
		\beq
		{\bf S}_M = \left( \frac{2}{3  }{\bf 1}-\mathbb{I} \right)
		\:\:\:\: . 
		\label{ct3.2}
		\eneq
		\noindent
		Following the derivation of Section \ref{sec:Noninteracting} and relating to general 
		arguments based on the scattering matrix description of the junction (see, e.g., \cite{Benenti2017}), 
		we conclude that the WFL holds at the ${\bf M}$FP.
		
		In general,  however, when the junction is not connected to  Fermi liquid reservoirs, we infer from Eqs.\eqref{ct3.1} 
		and \eqref{ct3.2} that we cannot describe the ${\bf M}$FP, within the bosonization framework, in terms of an orthogonal splitting matrix. 
		We may instead still define a non orthogonal matrix 
		\beq
		\tilde{\bf \rho} = \frac{1}{3}\left( \frac{4}{3} {\bf 1} -\mathbb{I} \right)
		\:\:\:\: , 
		\label{ct3.3}
		\eneq
		\noindent
		to describe the linear relations between the chiral electric current operators at the junction. The formalism of Appendix \ref{Renormalization} 
		cannot be exploited, as it is based on the possibility of expressing the currents in bosonic language. The non-orthogonality of the matrix in
		Eq.\eqref{ct3.3}, however, implies that it is not possible to describe the ${\bf M}$FP in terms of conformal boundary conditions  on the chiral bosonic fields. 
		
		In order to  capture  the main behavior of the junction connected to Fermi liquid   reservoirs, we refer to   Eqs. \eqref{Gpert} and \eqref{Kpert}. The conductances both receive 
		corrections proportional to $ {\cal J} ( D = 2 \pi k_B T )$.  On connecting the junction to the reservoirs, there is a 
		crossover in the scaling properties as soon as $\frac{\beta}{2 \pi} \sim \frac{\ell}{\pi u}$. At lower temperatures, we expect the scaling behavior for
		$g = g_{{\rm Res}} = 1$: the boundary interaction becomes, therefore, marginal and can change the ${\bf S}$-matrix of the junction.
		The ${\bf M}$FP, as well as the $\chi_\pm$FPs, are just specific points over the manifold spanned along the above marginal deformation. Accordingly, 
		they can all be equivalently described in terms of a single-particle ${\bf S}$ matrix. At each point of 
		that manifold the WFL holds \cite{Benenti2017}. 
		
		At the $D^2$FPs, the WFL is instead violated, with the Lorenz ratio computed in Eq. \eqref{ren.12}.
		Employing the formalism of Appendix \ref{General:expression}, we may write the scaling functions for the conductance tensors, as 
		well as for the renormalization factor of the Lorenz ratio, once we know the leading boundary perturbation allowed by the symmetries
		of the FP. For the $N=3$ junction, it was shown in \cite{Oshikawa2006} that the 
		leading boundary perturbation  corresponds  to a linear combination of boundary operators
		with scaling dimension $\Delta_P=\frac{g}{3}$ at the $D_P$ FPs and $\Delta_N=\frac{g}{9}$ at the $D_N$ FPs.
		
		From scaling arguments and symmetry considerations, we expect for the conductance tensors in the vicinity of  the FPs the general expressions 
		\begin{eqnarray}\label{ct3.4}
			\mathbb{G} (T) & = & \mathbb{G}^* - \frac{e^2\tilde{h}^2( 2 \pi k_B T)}{2 \pi} \left( 3 \mathbb{I} - {\bf 1} \right) \Phi_{{\rm el} } ( g )   \\
			\mathbb{K} (T) & = & \mathbb{K}^* - \frac{ \pi k_B^2 \tilde{h}^2( 2 \pi k_B T) T}{6} 
			\left( 3 \mathbb{I} - {\bf 1} \right) \Phi_{{\rm th}} ( g )\nonumber
			\:\:\:\: , 
		\end{eqnarray}
		\noindent
		with the dimensionless effective coupling  \mbox{$ \tilde{h}( D )=  \frac{ h}{D_0} \left( \frac{ D}{D_0}  \right)^{ - 1 + \Delta_{D_P (D_N)}} $}  
		and  $\mathbb{G}^*$ ,   $\mathbb{K}^*$ the FP conductance tensors in Eqs.\eqref{ren.11}. Eqs.\eqref{ct3.4},  with $\Phi_{{\rm el}/{\rm th}} ( g )=\Phi_{{\rm el}/{\rm th} ;D_P (D_N) } ( g )$, can be readily recovered using the formulas of Appendix \ref{General:expression}.
		According to Eqs.\eqref{ct3.4}, we find a corresponding renormalization of the Lorenz ratio given by 
		\begin{equation}
			{\cal L} ( T ) \approx    L_0  \left\{ \frac{2}{3} +  
			[ \Phi_{{\rm th} } ( g ) - \Phi_{{\rm el} } ( g ) ]\tilde{h}^2( 2 \pi k_B T) \right\}
			\:\: . 
			\label{ct4.4}
		\end{equation}
		\noindent
		As expected, when the FP is attractive, the finite-$T$ corrections to the FP conductance tensors, as well as to the Lorenz ratio, 
		scale to zero as $T \to 0$. 
		When, instead, the FP is unstable against a finite boundary coupling, we see from
		Eqs. \eqref{ct3.4} and \eqref{ct4.4} that the perturbative regime breaks down as soon as $\tilde{h}( 2 \pi k_B T ) \sim 1$. 
		Considering the case in which the junction is connected to the external reservoirs,  we expect that Eqs. \eqref{ct3.4} and \eqref{ct4.4}  cease 
		to be valid once again at a scale $ ( 2 \pi k_B T_c )^{-1}\sim \frac{\ell}{\pi u}$. At lower temperatures, $h$ is traded for the 
		running coupling extracted at the scale $T_c$ and with $\Delta_P = \frac{1}{3} , \Delta_N = \frac{1}{9}$. 
		We summarize the phase diagram for the connected $N=3$ junctions  in Fig.\ref{con_ju}. We see that the system always flows back to 
		the fixed point manifold described in terms of a single-particle ${\bf S}$ matrix. Therefore, we always recover the WFL  at low enough $T$.

		\begin{figure}
			\includegraphics*[width=0.9 \columnwidth]{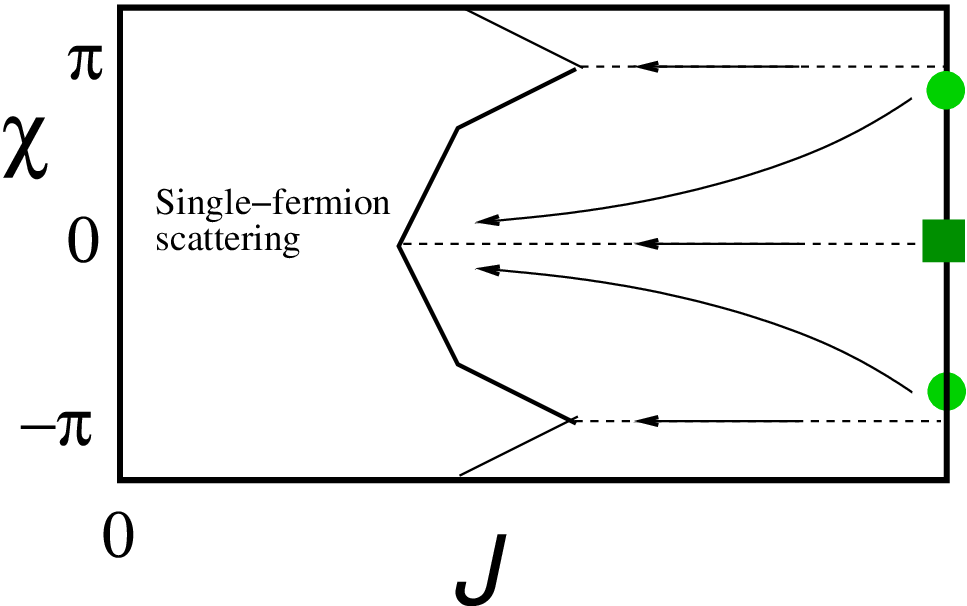}
			\caption{Sketch of the RG trajectories  in the ${\cal J} - \chi$ plane for the $N=3$ junction connected to Fermi liquid   reservoirs. 
				The $D_P$ and the $D_N$ FPs (light green dots and dark green squares, respectively) are both repulsive. The RG trajectories
				always flow back toward the fixed manifold each point of which is described in terms of single-fermion
				scattering processes. Accordingly, the WFL  is expected to be preserved as $T \to 0$, regardless
				of the initial values of the junction parameters.} 
			\label{con_ju}
		\end{figure}
		
		In the RG flow of the TKM, two nontrivial fixed point appear. At the $D$FP, for $N=3$, the leading boundary perturbation is a linear combination of operators with scaling dimension 
		$\Delta_D = \frac{4g}{3}$ \cite{Beri2012,Altland2013,Herviou2016,Giuliano2010}. At the $\hat{D}$FP, only reached along the fixed lines $\chi = \pm \pi$, the leading boundary
		perturbation has scaling dimension $\Delta_{\hat{D}} = \frac{4g}{9}$ \cite{Yi1998,Giuliano2010}. 
		In the vicinity of the FPs the conductance tensors have the form \eqref{ct3.4},
		with  $\tilde{h}= h \left( \frac{\beta}{2 \pi} \right)^{ 1 - \Delta_{D (\hat{D})}}$ and $\Phi_{{\rm el}/{\rm th} ; D (\hat{D}) } ( g )$ nonuniversal functions of $g$.
		$G ( T ) $ and $K ( T)$ flow to their fixed-point value with a leading, finite-$T$ correction scaling as $T^{2\left(\Delta_D-1\right)}$ and as $T^{2\Delta_D-1}$, respectively \cite{Us_short2021}. 
		The Lorentz ratio is corrected as in \eqref{ct4.4}, but this time in the {\it connected} junction and with \mbox{$\tilde{h}=h\left(\frac{\beta}{2\pi}\right)^{-\frac{1}{3}}$}.
		In Fig.\ref{con_ju_tk} we summarize the boundary RG flows of the TKM, to be compared to the one in Fig.\ref{con_ju}.
		The stable phase is completely different from the one that emerges in the simple $N=3$ connected junction.
		This is due to the peculiar scattering dynamics at the junction, tightly related to the emergence of the MZMs at the central island
		and to the hybridization between the MZMs and the Klein factors used in the bosonization of the leads \cite{Beri2013,giuliano2019}, 
		which in turn washes out the effect of the Klein factors that destabilize the $D^{N-1}$FP in a junction of normal wires \cite{Oshikawa2006,Giuliano2009,Giuliano2010}. 
		These considerations  eventually lead to the proposal of synoptically looking at the charge and at the thermal transport properties of the junction, as an alternative mean to characterize the MZMs at the island \cite{Us_short2021}.
		
		Before concluding this Section, it is worth stressing that, in the specific context of the Kondo effect,  a violation of the WFL  has been evidenced as $T$ is of the order of the Kondo temperature $T_K$  \cite{Costi2010}. 
		As we discuss in Appendix \ref{app:RGTKM}, in our specific case, $T_K$ is defined as the temperature scale at which  the perturbation theory breaks down and the running couplings become of order 1 \cite{Beri2012,Us_short2021}.
		Our result in Eq.\eqref{Gpert} is expected to apply only close to the DFP,  for $\mathcal{J}_K\left(2\pi k_B T\right)\ll1$, or, equivalently, $T\gg T_K$. Conversely, our results in Eqs. \eqref{ren.11} and \eqref{ct3.4} 
		hold near by the strongly coupled FP \cite{Us_short2021} and are therefore valid for  $T\ll T_K$.  For $g<1$, instead, the perturbative results are also reliable for $T\ll T_K$, provided the bare coupling strength is below the critical 
		value given in section \ref{ttkm}. In general, the full scaling curve (as a function of $T / T_K$) for both the charge and the thermal conductance has been numerically derived in \cite{Costi2010}, getting  two different values for $T_K$ from 
		the two scaling curves.   Our result in Eqs. \eqref{Gpert} and \eqref{Kpert} implies   that $ G(T)$ and $K(T)/T$ scale with the same function of $T$, 
		up to a factor $ L_0$.  In fact,  our perturbative derivation possibly misses higher-order contributions in ${\cal J}$ to the right-hand side of   Eqs. \eqref{Gpert} and \eqref{Kpert} which, in principle, might   render the scaling function associated to $G(T)$ and to $K(T) / T$ no longer different by only a factor $L_0$. Yet, while this is likely to affect the numerical estimate(s) of the Kondo temperature, 
		it is not expected to spoil our fixed points (and close to the fixed point) results.

		\begin{figure}
			\includegraphics*[width=0.9 \columnwidth]{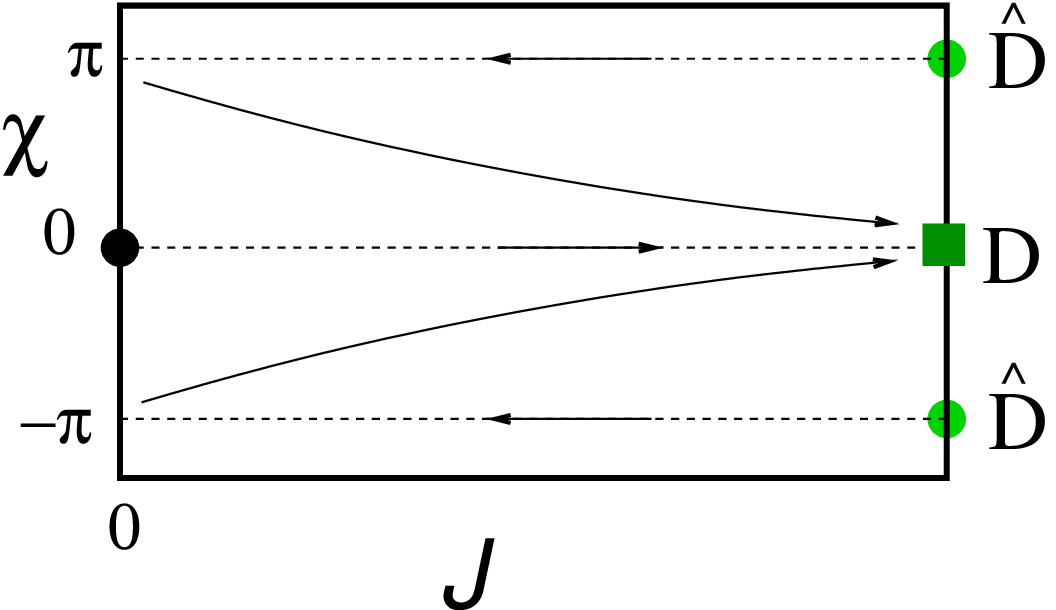}
			\caption{Sketch of the RG trajectories  in the ${\cal J} - \chi$ plane for the TKM connected to Fermi liquid reservoirs. 
				The $D$ and the $\hat{D}$ FPs (dark green squares and light green dots, respectively) are one attractive, the other(s) repulsive. The RG trajectories
				flow toward the $D$FP anywhere, except for $\chi = \pm \pi$.  Accordingly, at any $\chi \neq \pm \pi$, we 
				expect the Wiedemann-Franz to break down  as $T \to 0$, regardless
				of the initial values of the junction parameters.} 
			\label{con_ju_tk}
		\end{figure}
		\noindent

		\section{Conclusions}
		\label{sec:Conclusions}
		
		In this paper we determined the charge and the thermal conductance tensors  at various fixed points in the phase diagram of a junction of $N$ interacting  quantum wires. 
		We showed the direct connection between the onset of  Andreev reflection and/or crossed Andreev reflection processes and the violation of the Wiedemann-Franz law determined by the corresponding 
		"charge-heat separation" \cite{Kane1994,Kane1997,Benenti2017}. 
		In the specific case in which the total charge is conserved at the junction, we have shown 
		that the breakdown of the Wiedemann-Franz law  is directly related to the onset of multiparticle scattering processes and that it is different from the "trivial" breakdown determined by interactions.
		For $N=3$ wires, we have explicitly computed the Lorenz ratio for a wide class of boundary conditions.
		
		Among the possible mechanisms stabilizing a phase with multiparticle scattering at the central island, we have pointed out the role of the bulk
		interaction in the leads and explored the consequences of the coupling between isolated Majorana modes at the central island and the low-energy modes 
		in the leads. We have highlighted that, when connecting the junction to external, Fermi liquid reservoirs (as fairly common in transport experiments), 
		the former mechanism is deactivated, while the latter mechanism remains effective.
		In this paper and in \cite{Us_short2021}, we have explored the direct relation between the breakdown of the Wiedemann-Franz law  and the presence of Majorana zero modes  in the junction.
		
		The effectiveness of the combined analysis of the charge and heat transport properties of a junction of quantum wires  to unveil the relevant physics that 
		sets in at nontrivial fixed points  in the phase diagram of the system suggests extensions of our approach to, e.g., junctions of quantum spin chains 
		(where the charge current has to be substituted with a pertinent definition of the spin current) \cite{Tsvelik2013,Crampe2013,Buccheri2015,Giuliano2016a}, 
		or to junction of bosonic systems \cite{Tokuno2008}, such as cold atom condensates \cite{Buccheri2016}. We plan to go through this topic as a further extension of our work. 
		
		\vspace{0.5cm}
		
		{\bf Acknowledgements}: We thank M. Burrello for the valuable feedback. F.B. and R.E. were 
		funded by the Deutsche Forschungsgemeinschaft (DFG, German Research Foundation) under Germany's Excellence Strategy Cluster of Excellence Matter and Light for Quantum Computing (ML4Q) \mbox{EXC 2004/1  390534769} and Normalverfahren Projektnummer \mbox{EG 96-13/1} and under Projektnummer \mbox{277101999 - TRR 183} (project B04). 
		A. N. was financially supported by POR Calabria FESR-FSE 2014/2020 - Linea B) Azione 10.5.12,  grant no.~A.5.1.  D.G.   acknowledges financial support  from Italy's MIUR  PRIN project  TOP-SPIN (Grant No. PRIN 20177SL7HC).

		\appendix 
		
		\section{Lattice model for a junction of $N$ noninteracting quantum wires}
		\label{Noninteracting}
		
		In this Appendix we briefly review the derivation of the CCT  and of the HCT 
		in a lattice model of a junction of $N$ noninteracting quantum wires. In particular, we assume that,  
		as the temperature $T \to 0$, only single-particle 
		scattering processes take place at the central island: we allege a particle/hole entering from wire $j$ to undergo a  normal reflection 
		within the same wire as a backscattered particle/hole, 
		a normal transmission to wire $j'$ ($\neq j$) as a particle/hole, an  Andreev reflection 
		within the same wire as a hole/particle, or a  crossed Andreev reflection to wire $j'$ ($\neq j$) as 
		a hole/particle. To account for all these processes we employ a pertinently adapted lattice version of the 
		two-component Nambu spinor formalism of \cite{Blonder1982}. Regarding each lead as an $\ell$-site lattice 
		with hopping strength $J$ and chemical potential $\mu$, we write the lattice lead Hamiltonian as 
		$H_{{\rm Lat} , 0}=\sum_{ j = 1}^N \: H_{{\rm Lat} , 0,j}$, with
		\beq
		H_{{\rm Lat} , 0,j} = - J \sum_{r = 1}^{\ell - 1} 
		[ c_{ r , j}^\dagger c_{r+1, j } + c_{r+1 , j}^\dagger c_{r , j } ] - \mu \sum_{r =1}^\ell 
		c_{ j , r}^\dagger c_{j , r }
		\:\: , 
		\label{nonia.1}
		\eneq
		with $c_{r , j } , c_{ r , j}^\dagger$ being single-fermion annihilation/creation operators at
		site $r$ of lead $j$. Within our formalism, we write the wavefunction for  an incoming 
		particle from wire $j$, evaluated at site $r$ of wire $j'$ as  
		\begin{eqnarray}
			&& \left[ \begin{array}{c}
				u_{r ; \epsilon ; j'}^{(j , p)} \\   v_{r ; \epsilon ; j'}^{(j , p)}        
			\end{array} \right] 
			= \delta_{j , j'} \: \mathcal{N}_p \left[ \begin{array}{c}
				e^{ - i k_p (r-1 ) } + r_{j',j'} ( \epsilon )    e^{  i k_p(r-1 ) } \\
				a_{j',j'} ( \epsilon )    e^{  - i k_h(r-1 ) } 
			\end{array} \right] \nonumber \\
			&&\qquad\qquad + [ 1 - \delta_{j , j'} ] \:     \mathcal{N}_p  \left[ \begin{array}{c}
				t_{j,j'} ( \epsilon )    e^{  i k_p  (r-1 ) } \\
				c_{j,j'} ( \epsilon )    e^{  - i k_h (r-1 ) } 
			\end{array} \right]                              
			\;\;\;\; , 
			\label{anomalo.1}
		\end{eqnarray}
		\noindent
		with $r_{j,j} ( \epsilon )$, $a_{j,j} ( \epsilon)$, $t_{j,j'} ( \epsilon)$ , $c_{j,j'} ( \epsilon )$ respectively 
		corresponding to the  normal reflection amplitude within wire $j$, to the Andreev reflection amplitude within 
		wire $j$, to the normal transmission amplitude from wire $j$ to wire $j'$, and to the crossed Andreev reflection amplitude 
		from wire $j$ to wire $j'$. Also, we write the similar wavefunction for an  incoming hole  
		from lead $j$ as 
		\begin{eqnarray}
			&& \left[ \begin{array}{c}
				u_{r ; \epsilon ; j'}^{(j , h)} \\   v_{r ; \epsilon ; j'}^{(j , h)}        
			\end{array} \right] 
			= \delta_{j , j'} \: \mathcal{N}_h \left[ \begin{array}{c}
				\tilde{a}_{j',j'} ( \epsilon )    e^{  i k_p (r-1 ) } \\
				e^{ i k_h (r-1)} + \tilde{r}_{j',j'}  ( \epsilon )    e^{  - i k_h  (r-1 ) } 
			\end{array} \right] \nonumber \\
			&&\qquad\qquad + [ 1 - \delta_{j' , j} ] \:      \mathcal{N}_h  \left[ \begin{array}{c}
				\tilde{c}_{j,j'} ( \epsilon )    e^{  i k_p  (r-1 ) } \\
				\tilde{t}_{j,j'} ( \epsilon )    e^{  - i k_h  (r-1 ) } 
			\end{array} \right]                              
			\;\;\;\; , 
			\label{anomalo.2}
		\end{eqnarray}
		\noindent
		with the amplitudes $\tilde{r}_{j,j} ( \epsilon )$, $\tilde{a}_{j,j} ( \epsilon )$, $\tilde{t}_{j,j'} ( \epsilon )$ and 
		$\tilde{c}_{j,j'} ( \epsilon )$ having the same meaning as those in Eq.\eqref{anomalo.1}. 
		The scalars $\mathcal{N}_{p , h}$ are normalization constants and the momenta $k_{p,h}$ are defined as a function of the energy by the relations
		$\epsilon  =   - 2 J \cos ( k_p )  - \mu = 2J \cos ( k_h ) +  \mu.$
		Denoting with   $ \{ c_{\epsilon , p , j} , c_{\epsilon , h , j} \}$   the single-fermion annihilation 
		operators in the state corresponding to a particle/hole entering  the central island from 
		lead $j$ at energy $\epsilon$, we write the real-space, lattice single-fermion operators as 
		\beq
		c_{ r , j } = \sum_{\epsilon > 0} \: \sum_{ j' = 1}^N \; \sum_{ u = p,h} \: 
		\{ u_{r ; \epsilon ; j }^{(j' , u)} c_{\epsilon , u , j'}  + [ v_{r ; \epsilon ; j }^{(j' , u)} ]^* c_{\epsilon , u , j'}^\dagger \}
		\:\:\:\: . 
		\label{anomalo.3}
		\eneq
		Expanding $c_{r , j }$ by retaining only the low-energy, long-wavelength excitations around the 
		Fermi momenta $\pm k_F = \pm {\rm arccos} \left( - \frac{\mu}{2J} \right)$, we can recast it in the form 
		\beq
		c_{ r , j } \approx e^{ i k_F  r} \psi_{R , j } ( x_r ) + e^{ - i k_F r} \psi_{L , j } ( x_r )
		\:\:\:\:,
		\label{anomalo.3.1}
		\eneq
		with $x_r = ar$ and $a$ being the lattice step (which we set to unity elsewhere in the paper). 
		$\psi_{R , j  } ( x ) , \psi_{L , j  } ( x)$ are the chiral fields that we used throughout the 
		derivation of Section \ref{sec:Noninteracting}, where we resorted to the continuum variable 
		framework. Here, instead, we keep using the lattice formalism, in which 
		the current operators in lead $j $ are given by 
		
		\begin{eqnarray} \label{anomalo.4}
			j_{{\rm el} , r , j } &=& - i e J  c_{r , j }^\dagger c_{ r + 1 , j } + {\rm h.c.}  \\
			j_{{\rm th} , r , j  } &=&  i J^2 c_{r - 1 , j }^\dagger c_{ r + 1 , j }+\frac{i \mu   J}{2} 
			c_{r , j }^\dagger \left( c_{ r + 1 , j } -c_{r -1 , j }\right)+ {\rm h.c.}  \nonumber
		\end{eqnarray}
		\noindent
		In order to compute the average values of the  operators in Eqs.\eqref{anomalo.4}, we assume 
		that each lead $j $ is connected to a thermal reservoir at voltage bias $V_j$ and at temperature $T_j = T + \delta T_j$.
		Collecting the contributions all together, we obtain the expectation value of the electric current 
		\begin{widetext}
			\begin{eqnarray}
				I_{{\rm el} , j } &=& \langle j_{{\rm el} , j} ( x , t ) \rangle =
				e \sum_\epsilon \sum_{ j' = 1}^N \Biggl\{  v_p \left[T_{j ,j'} ( \epsilon )-\delta_{j,j'}\right] f^{(p)}_{j'} ( \epsilon ) +  v_p  A_{j ,j'} ( \epsilon ) [ 1 - f^{(p)}_{j'} ( \epsilon ) ] \nonumber \\
				&&\qquad\qquad + v_h  \tilde{A}_{j ,j'} ( \epsilon ) f^{(h)}_{j'} ( \epsilon ) +  v_h \left[\tilde{T}_{j ,j'} ( \epsilon )-\delta_{j,j'}\right] [ 1 - f^{(h)}_{j'} ( \epsilon ) ] \Biggr\}
				\:\:\:\: , 
				\label{anomalo.5_V0}
			\end{eqnarray}
			\noindent
			with $v_{p/h} = v_{p/h} ( \epsilon ) = 2 J \sin ( k_{p/h} ( \epsilon ))$. Similarly, the 
			thermal current is
			\begin{eqnarray}
				I_{{\rm th} , j } &=&  \langle j_{{\rm th} , j} ( x , t ) \rangle =  \sum_\epsilon ( \epsilon - \mu)  \sum_{ j'  = 1}^N \Biggl\{  v_p    \left[T_{j ,j'} ( \epsilon )-\delta_{j,j'}\right]f^{(p)}_{j'} ( \epsilon )
				-  v_p   A_{j ,j'} ( \epsilon ) \left[ 1 - f^{(p)}_{j'} ( \epsilon ) \right] \nonumber \\
				&& + v_h   \tilde{A}_{j ,j'} ( \epsilon ) f^{(h)}_{j'} ( \epsilon ) 
				- v_h \left[\tilde{T}_{j ,j'} ( \epsilon )-\delta_{j,j'}\right] 
				[ 1 - f^{(h)}_{j'} ( \epsilon ) ] \Biggr\}
				\:\: , 
				\label{anomalo.6_V0}
			\end{eqnarray}
			\noindent
		\end{widetext}
		with the scattering coefficients given in Section \ref{sec:Noninteracting}
		and the Fermi distribution functions for particles and holes respectively given by Eq.\eqref{noni.4} in the main text. 
		In the large-$\ell$ limit, we trade the sum over the energies for integrals and introduce the density of states around 
		the Fermi energy $\rho_0$. Linearizing the Fermi distribution in the voltage and temperature biases 
		\begin{eqnarray}
			f^{(p)}_j ( \epsilon ) &\approx& f ( \epsilon  ) + \left[  \left(   \frac{\epsilon-\mu}{k_B T^2} \right)  \delta T_j   
			- \frac{ e V_j}{k_B T} \right]  \partial_\epsilon f ( \epsilon - \mu ) \nonumber \\
			f^{(h)}_j ( \epsilon ) &\approx& f ( \epsilon ) + \left[ \left(   \frac{\epsilon-\mu}{k_B T^2}  \right)  \delta T_j   
			+ \frac{ e V_j}{k_B T} \right]  \partial_\epsilon f ( \epsilon - \mu ) \;\; ,  \nonumber \\
			\label{anomalo.9}
		\end{eqnarray}
		and employing the unitarity of the extended ${\bf S}$-matrix, as well as the Sommerfeld expansion for the resulting integrals at temperatures $k_B T\ll \mu$, we obtain
		\begin{eqnarray}
			I_{{\rm el} , j} &=& - \frac{e^2}{2 \pi} 
			\; \sum_{ j' = 1}^N \left[ \delta_{j,j'} +C_{j,j'}  - T_{j,j'}\right] V_{j'} \nonumber \\
			I_{{\rm th} , j} &=&   \frac{\pi k_B^2 T }{6}
			\; \sum_{ j' = 1}^N \Biggl[ C_{j,j'} + T_{j,j'}-\delta_{j,j'}  \Biggr]  \delta T_{j'} 
			\:\:\:\: , 
			\label{anomalo.11}
		\end{eqnarray}
		\noindent
		with the dependence on $\epsilon$ in the scattering coefficients dropped to mean that all of them are computed at $\epsilon = \mu$. From Eqs.\eqref{anomalo.11}, cast in the form
		\begin{equation}
			I_{{\rm el} , j}= \sum_{j' = 1}^N G_{j,j'} V_{j'} \;,\;
			I_{{\rm th} , j} = \sum_{j' = 1}^N  K_{j,j'} \delta T_{j'} 
			\:\: ,
			\label{noni.5}
		\end{equation}
		we obtain Eq.\eqref{noni.6} in  the main text.
		As a side remark, we point out that throughout the paper we can safely compute the heat current by averaging the energy current 
		operator, rather than the heat itself. Since the two operators differ by $\sum_{j=1}^N V_j I_{{\rm el} , j}$, 
		we see that it is of second order in the biases $V_j , \delta T_j$. Thus, we can safely neglect it within linear response theory.
		
		\section{Electric and thermal conductance for a ballistic interacting single quantum wire}
		\label{ballistic_single}
		
		In this Appendix we review the calculation of the electric and of the thermal conductance for 
		a single interacting quantum wire connected to two reservoirs kept at different voltages and 
		temperatures. Besides reviewing well-known results \cite{Kane1992,Kane1992b,Kane1996,Kane1997},   we 
		set up our  formal approach to computing the CCT and the HCT. 
		
		\begin{figure}
			\center
			\includegraphics*[width=1. \columnwidth]{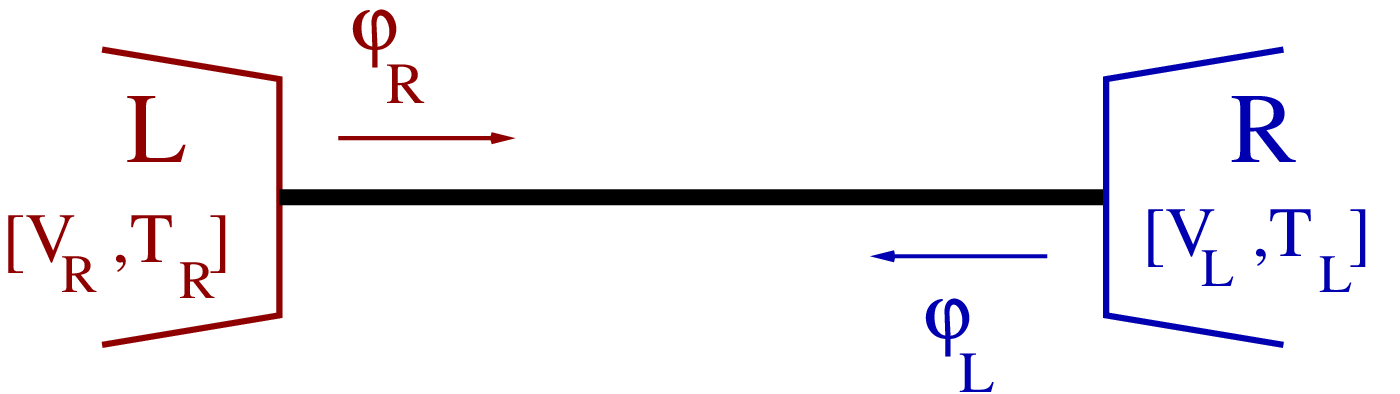}
			\caption{Sketch of a single, interacting quantum wire connected to a left-hand reservoir, which 
				injects into the system right-handed modes at voltage bias $V_R$ and at temperature $T_R$ (colored in  
				red), and to a right-hand reservoir that injects left-handed modes at voltage bias $V_L$ and at temperature $T_L$ (colored 
				in blue).  } 
			\label{singlewire}
		\end{figure}
		\noindent
		In Fig.\ref{singlewire} we sketch the wire connected to a left-hand reservoir, which 
		injects into the system right-handed modes at voltage bias $V_R$ and at temperature $T_R$, and to
		a right-hand reservoir that injects left-handed modes at voltage bias $V_L$ and at temperature $T_L$. 
		To describe the wire, we employ  a minimal model for the corresponding  Hamiltonian in fermionic coordinates, $H_{\rm Fer}$, 
		which is given by  
		\begin{eqnarray}
			H_{\rm Fer} &=& - i v \int_0^\ell  d x  \{\psi_R^\dagger ( x ) \partial_x \psi_R ( x ) 
			- \psi_L^\dagger ( x ) \partial_x \psi_L ( x ) \} \nonumber \\
			&+& 2 V  \int_{0}^\ell d x   \: \psi_R^\dagger ( x ) \psi_R ( x ) \psi_L^\dagger ( x ) \psi_L (  x ) 
			\:\: , 
			\label{lut.1}
		\end{eqnarray}
		\noindent
		with $V$ being the bulk interaction strength. Along the bosonization 
		approach, we introduce two chiral bosonic fields $\varphi_R ( x , t ) = \varphi_R ( t_-  )$
		and $\varphi_L ( x , t ) = \varphi_L ( t_+  )$. Their dynamics is governed by the Hamiltonian 
		\begin{eqnarray}
			H_{{\rm Bos} , 0 } &=& u \: \int_0^\ell \: d x \: \{ ( \partial_x \varphi_R ( x ))^2 + 
			( \partial_x \varphi_L ( x ))^2 \} \nonumber \\ 
			&\equiv& H_{{\rm Bos} , 0 , R} + H_{{\rm Bos} , 0 , L}
			\:\:\:\: ,
			\label{lut.2}
		\end{eqnarray}
		\noindent
		with the plasmon velocity $ u = v \sqrt{ 1 - \frac{V^2}{ v^2 \pi^2}}$. 
		The fermion operators are represented via
		\begin{eqnarray}
			\psi_R ( x ) &=& \Gamma \: e^{ \sqrt{\pi} \left[ \left( \frac{1}{\sqrt{g}} + \sqrt{g} \right)  \varphi_R ( x ) + 
				\left( \frac{1}{\sqrt{g}}-  \sqrt{g} \right)  \varphi_L ( x ) \right] } \nonumber \\
			\psi_L  ( x ) &=& \Gamma \: e^{ \sqrt{\pi} \left[ \left( \frac{1}{\sqrt{g}} - \sqrt{g} \right)  \varphi_R ( x ) + 
				\left( \frac{1}{\sqrt{g}}+  \sqrt{g} \right)  \varphi_L ( x ) \right] }
			\:\:,
			\label{lut.4}
		\end{eqnarray}
		in which the Klein factors $\Gamma$ can  be safely discarded, when bosonizing a single QW.  
		Finally, we rewrite the densities and the current operators in terms of 
		the chiral fields as 
		\begin{eqnarray}
			\rho_{\rm el} ( x , t ) &=&  e \sqrt{\frac{g}{\pi}}\: \{ \partial_x \varphi_R ( x , t ) - \partial_x \varphi_L ( x , t ) \} 
			\nonumber \\
			\rho_{\rm th} ( x , t ) &=& u \: \{( \partial_x \varphi_R ( x , t ))^2 + ( \partial_x \varphi_L ( x , t ))^2 \} 
			\:\:\:\: ,
			\label{lut.5}
		\end{eqnarray}
		\noindent
		and 
		
		\begin{eqnarray}
			j_{\rm el} ( x , t ) &=& e u \sqrt{\frac{g}{\pi}} \; 
			\{\partial_x \varphi_R ( x , t ) + \partial_x \varphi_L ( x , t ) \}  \nonumber \\
			j_{\rm th} ( x , t ) &=&  u^2 \: 
			\{ (\partial_x \varphi_R ( x  , t ))^2 - ( \partial_x \varphi_L ( x , t ))^2 \}  
			\:\:\:\: . 
			\label{lut.6}
		\end{eqnarray}
		\noindent
		In order to take into account that modes with opposite chiralities are 
		``shot in'' from the reservoirs at different voltage bias (which couple with 
		the corresponding charge density operators), we modify the wire Hamiltonian 
		by adding a voltage dependent ``source'' term. As a result, we 
		obtain

		\begin{eqnarray}
			H_{\rm Bos}  [ V_L , V_R ] &=& u \int_0^\ell \: d x \: \{ ( \partial_x \varphi_R ( x ))^2 + ( \partial_x \varphi_L ( x ))^2 \} \nonumber \\
			&-& e \sqrt{\frac{g}{\pi}} \: \int_0^\ell \: d x \: \{ V_R \partial_x \varphi_R ( x ) - V_L \partial_x \varphi_L ( x ) \} \nonumber \\
			&\equiv& H_{{\rm Bos} , L} [ V_L ] +  H_{{\rm Bos} , R} [ V_R ]
			\:\:\:\: . 
			\label{lut.7}
		\end{eqnarray}
		\noindent
		Chiral modes are shot in from the reservoirs at different temperatures, as well. In order to account for 
		the different temperature of the opposite chirality modes,  given a generic operator ${\cal O}$ that is 
		a functional of $\varphi_R ( x ) , \varphi_L ( x )$, we compute its thermal average $\langle {\cal O} \rangle$
		according to

		\beq
		\langle {\cal O} \rangle = \frac{ {\rm Tr} [{\cal O}  e^{ - \beta_L  H_{{\rm Bos} , L} [ V_L ]
				- \beta_R  H_{{\rm Bos} , R} [ V_R ] } ]  }{{\cal Z}_L [ V_L , \beta_L ] {\cal Z}_R
			[V_R , \beta_R ] }
		\:\:\:\: , 
		\label{lut.8}
		\eneq
		\noindent
		with $\beta_{R , L } = ( k_B T_{R , L} )^{-1}$ and the partition functions for the 
		chiral modes being given by 
		\beq
		{\cal Z}_{L , R} [ V_{L , R } , \beta_{ L , R } ] = {\rm Tr} [ e^{ - \beta_{ L , R} H_{{\rm Bos} , L , R} } ] 
		\:\:\:\: . 
		\label{lut.9}
		\eneq
		\noindent
		It is customary to define the shifted chiral fields $\bar{\varphi}_R (x) , \bar{\varphi}_L (x)$, according to 
		\begin{eqnarray}
			\partial_x \bar{\varphi}_R ( x )  &=& \partial_x \varphi_R ( x ) - \frac{e}{2u} \sqrt{\frac{g}{\pi}} V_R \nonumber \\
			\partial_x \bar{\varphi}_L ( x )  &=& \partial_x \varphi_L ( x ) + \frac{e}{2u} \sqrt{\frac{g}{\pi}} V_L
			\:\:\:\: . 
			\label{lut.10}
		\end{eqnarray}
		\noindent
		Taking advantage of the system homogeneity, we compute the average value of the 
		charge- and of the thermal-current, $I_e$ and $I_h$, as 
		\beq
		I_{{\rm el}/{\rm th}} = \frac{1}{\ell} \: \int_0^\ell \: d x \: \langle j_{{\rm el}/{\rm th}} ( x ) \rangle
		\:\: . 
		\label{lut.13}
		\eneq
		\noindent
		As a result, retaining only linear contributions in the voltage/temperature bias and using the relation
		\beq
		u^2\int_0^\ell d x 
		\left\langle  ( \partial_x \bar{\varphi}_{L/R} ( x ))^2 \right\rangle
		= \frac{1}{u} \partial_{\beta_{L/R} } \ln {\cal Z}_{L , R}\;,
		\eneq
		we obtain
		\beq
		I_{\rm el}=\frac{e^2 g}{2 \pi} ( V_R - V_L )\quad,\quad I_{\rm th}=\frac{\pi}{12 \beta_R^2} - \frac{\pi}{12 \beta_L^2}\: . 
		\eneq
		Setting $\Delta V = V_R - V_L$ and $T_R = T + \frac{\Delta T}{2}, T_R = T - \frac{\Delta T}{2}$, the charge and the thermal conductance follow directly 
		\begin{eqnarray}
			G = \frac{I_{\rm el}}{\Delta V} = \frac{g e^2}{2 \pi} &\quad,\quad&
			K = \frac{I_{\rm th}}{\Delta T} = \frac{\pi k_B^2 T}{6}
			\:\: , 
			\label{lut.15}
		\end{eqnarray}
		as derived in \cite{Kane1996}. In this paper, it was  observed that Eq.\eqref{lut.15} implies  the violation of the WFL, as
		evidenced by the renormalization of the Lorenz ratio  to $\mathcal{L}= {L_0}/{g}$. This is directly related to the peculiar nature of the electronic 
		interaction in  one-dimensional systems,  that typically   cannot be accounted for within the Fermi liquid  picture.
		Nevertheless, we also note that, since the renormalization of $L_0$ is associated to the interaction-dependent renormalization of  $G$
		in a Luttinger liquid, we expect that it is washed out as soon as the 1d interacting system is connected to  external Fermi liquid reservoirs,
		in perfect analogy with what happens to the electric conductance of a Luttinger liquid   connected to noninteracting reservoirs \cite{Ponomarenko1995,Safi1995,Maslov1995,Oshikawa2006}. 
		In Appendix \ref{Renormalization} we show that this is the case. As we discuss in the main text, only at a junction where 
		multi-particle scattering processes take place even in the noninteracting limit, one recovers a "genuine" violation of the WFL,
		that survives the presence of Fermi liquid  reservoirs attached to the interacting QWs.

		\section{Renormalization of the conductance tensors at a junction of Luttinger liquids connected to Fermi liquid reservoirs}
		\label{Renormalization}
		
		In this Appendix we derive the  renormalization of the conductance tensor of a junction of interacting QWs connected to external Fermi liquid reservoirs. For the charge conductance, this is a well-known effect, due to the 
		fact that, over long enough length scales and/or at low enough energies, the collective modes of the Luttinger
		liquid are determined by the parameters of the external leads  
		\cite{Ponomarenko1995,Safi1995,Maslov1995,Oshikawa2006}. We extend the analysis to the thermal 
		conductance tensor, so to  determine the effects of the Fermi liquid reservoirs directly on the Lorenz ratio.
		
		Following  \cite{Ponomarenko1995,Safi1995,Maslov1995,Oshikawa2006}, 
		we model the reservoirs in terms of a discontinuity of the Luttinger parameter in each wire at $x=\ell$. Specifically,
		we set 
		
		\begin{eqnarray}
			g ( x ) &=& \Biggl\{ \begin{array}{l} g \;\;\; , \;\; 0 < x < \ell \\
				g_{\rm Res} \;\;\; , \;\; \ell \leq x 
			\end{array} \nonumber \\
			u  ( x ) &=& \Biggl\{ \begin{array}{l} u \;\;\; , \;\; 0 < x < \ell \\
				u_{\rm Res} \;\;\; , \;\;\ell \leq x \end{array}
			\:\: ,
			\label{jul.16}
		\end{eqnarray}
		\noindent
		In the following, we denote with $\varphi_{R/L , {\rm Res} , j }$ and with $\varphi_{R /L , j}$ the chiral bosonic fields 
		within the Fermi liquid reservoirs and within the interacting QWs, respectively. At the interface at $x=\ell$, the continuity of 
		the fields $\phi_j ( x )$, as well as of the charge current density operators,  
		implies the linear relations \cite{Hou2012} 
		
		\begin{equation} \label{eqinterface}
			\left[ \begin{array}{c} \varphi_{R , {\rm Res} , j} \left( \ell \right)
				\\
				\varphi_{L , {\rm Res} , j} \left( \ell \right) \end{array} \right] 
			= {\bf m} \cdot
			\left[ \begin{array}{c} \varphi_{R ,  j} \left( \ell \right)
				\\
				\varphi_{L , j} \left( \ell \right) \end{array} \right]  
		\end{equation}
		between the chiral fields of the system and of the reservoirs at the contact point.  The orthogonal interface transfer matrix ${\bf m}$ 
		is given by 
		
		\beq
		{\bf m} = \frac{1}{2 \sqrt{g g_{\rm Res}}}\left[ \begin{array}{cc}
			g_{\rm Res}+ g & g_{\rm Res}- g  \\
			g_{\rm Res}- g & g_{\rm Res}+ g
		\end{array} \right]
		\equiv \left[ \begin{array}{cc}
			m_{R R} & m_{R L} \\
			m_{L R} & m_{R R} 
		\end{array}
		\right]                          
		\:\: . \label{juf.11}
		\eneq
		The reservoirs inject chiral $L$ modes, whose dynamics is governed by the Luttinger Hamiltonian with parameter $g_{\rm Res}$.
		Having in mind a setup in which the current measurement is performed via the reservoir $j$ (as in \cite{Us_short2021}), we
		need to express the outgoing modes in that reservoir in terms of the incoming ones (in any reservoir). This task is performed by a pertinent 
		splitting matrix $\hat{\rho}$, defined via
		
		\beq
		\varphi_{R , {\rm Res} , j}\left(t-\frac{x}{u_{\rm Res}}\right)=\sum_{j'=1}^3 \hat\rho_{j,j'}\varphi_{L , {\rm Res} , j'}\left(t+\frac{x}{u_{\rm Res}}\right)
		\eneq
		\noindent
		Combining the interface transfer matrix Eq.\eqref{juf.11} with the junction splitting matrix Eq.\eqref{eq:rho}, 
		we obtain 
		
		\beq\label{rhohat}
		\hat\rho_{j,j'} = \sum_{j'' = 1}^N \; [ M_R ]_{j , j''} \: [ M_L^{-1} ]_{j'' , j'}\;\:\: ,
		\eneq
		\noindent
		with 
		
		\begin{eqnarray}
			\left[M_R\right]_{j,j'} &=& m_{RR} \rho_{j,j'} + m_{RL} \delta_{j,j'} \;, \nonumber\\
			\left[M_L\right]_{j,j'}  &=& m_{LR} \rho_{j,j'} + m_{LL} \delta_{j,j'} \; .
			\label{melem}
		\end{eqnarray}
		\noindent
		We therefore  conclude that the effect of the reservoirs is taken into account by substituting the 
		splitting matrix $\rho$  with $\hat{\rho}$  defined in \eqref{rhohat}.
		Substituting the parametrization Eqs. \eqref{para.3} and \eqref{para.4} for the charge-conserving junction, one checks that, for 
		the $N=3$, $\mathbb{Z}_3$ invariant junction,  the resulting splitting matrix is obtained by simply substituting $g$ with $g_{\rm Res}$ 
		in the splitting matrix of the junction disconnected from the reservoirs, that is
		\beq
		\hat\rho(\vartheta) = \left.\rho(\vartheta)\right|_{g\to g_{\rm Res}}
		\:\:\: .
		\label{subs}
		\eneq
		\noindent
		At variance, when charge conservation holds but $\mathbb{Z}_3$ symmetry is broken,  Eq.\eqref{para.9},  
		the splitting matrices  $\rho_B$  are independent of  $g$ and of $g_{\rm Res}$, so one trivially obtains the same result
		with, or without, connecting the junction to Fermi liquid  reservoirs.

		\section{Green-Keldysh functions of bosonic operators}
		\label{vertex_keldysh}
		
		In this appendix, we  review  the Keldysh-Green functions involving chiral 
		bosonic vertex operators, in the time, as well as in the frequency domain. 
		
		We start with the Keldysh path-ordered correlation function of the chiral fields
		\begin{eqnarray}  \label{ft.1}
			&& g_{\eta_1 \eta_2 ; j}^{-\nu}  \left( t_1 - t_2 + \frac{x_1 - x_2}{u} \right) = \\
			&& \langle {\cal T}_K \: e^{ i \sqrt{4 \pi \nu} \varphi_{j,\eta_1} ( t_1 + x_1/u  ; \eta_1)  }  e^{ - i \sqrt{4 \pi \nu} \varphi_{j,\eta_2}( t_2  + x_2/u ; \eta_2 ) }  \rangle
			\:\:, \nonumber
		\end{eqnarray}
		\noindent
		with the Keldysh indices $\eta_{ 1 , 2 } = \pm 1$ and ${\cal T}_K$ denoting the ordering operator along 
		the Keldysh path. The Green's functions are \cite{Campagnano2016}
		\begin{eqnarray}
			g_{++ , j}^{-\nu} \left( t + \frac{x}{u} \right) &=& \left\{ \frac{\beta_j}{\pi} \sin \left[ \frac{\pi}{\beta_j} \left( i  \left( t + \frac{x}{u} \right) {\rm sgn}  ( t ) + \tau_c \right) \right] \right\}^{-\nu} \nonumber \\
			g_{-+ , j}^{-\nu}  \left( t + \frac{x}{u} \right) &=& \left\{ \frac{\beta_j}{\pi} \: \sin \left[ \frac{\pi}{\beta_j} \left( i \left( t + \frac{x}{u} \right) + \tau_c \right) \right] \right\}^{-\nu} 
			\label{gppgmp}
		\end{eqnarray}
		with \mbox{$g_{-\eta_1-\eta_2 , j}^{-\nu}\left(t+\frac{x}{u}\right) = \left[g_{\eta_1\eta_2 , j}^{-\nu}\left(-t-\frac{x}{u}\right)\right]$} and \mbox{$\tau_c \sim D_0^{-1}$}.
		Their Fourier transforms
		\beq
		\tilde{g}_{\eta_1\eta_2,j}^{-\nu} ( \omega ) = \int d t\, e^{ i \omega t}  g_{\eta_1\eta_2, j}^{-\nu} \left( t  \right)
		\eneq
		are derived in detail in \cite{Campagnano2016}. Here we quote the result
		\beq\label{eq:ftg}
		\tilde{g}_{\eta_1\eta_2,j}^{-\nu}(\omega ) =
		c_{\eta_1\eta_2,j}^{-\nu}(\omega) d_j^{-\nu}(\omega) \;,
		\eneq
		with, denoting by $\Gamma ( z )$ the Euler's $\Gamma$ function,
		\beq \label{ft.6}
		d_j^{-\nu}(\omega) = \frac{ \beta_j^{1- \nu}}{\left({2 \pi} \right)^{1- \nu}\Gamma ( \nu ) }  \left| \Gamma \left( \frac{\nu}{2} + i \frac{\beta_j \omega}{2 \pi} \right) \right|^2 \;,
		\eneq
		\beq \label{ft.7}
		c_{\pm\pm,j}^{-\nu} ( \omega ) = \frac{2\cosh \frac{\beta_j \omega}{2}  }{1+e^{\pm i\pi \nu}}\quad,\quad
		c_{\mp\pm,j}^{-\nu} ( \omega ) =  e^{ \pm\frac{\beta_j \omega}{2}} 
		\:\: . 
		\eneq
		Eq. \eqref{ft.1} is easily generalized to a multipoint correlation function. In particular, we need the multiple contraction
		\begin{equation}
			g_{\{\eta_n\}_n; j}^{\{\alpha_n\}_n; N ; - \nu } \left( \{t_n\}_n \right) = \delta_{\sum_n\alpha_n,0}\prod_{l<m}
			g_{\eta_l,\eta_m; j}^{ -\alpha_l\alpha_m \nu }\left( t_l-t_m \right)
			\:\: . 
			\label{ft.4}
		\end{equation}
		It is also convenient to define the functions 
		\beq
		\xi_{\pm , j } \left(  u t + x \right) =  \frac{\pi}{i\beta_j} \coth \left[ \frac{\pi}{\beta_j} \left(  t + \frac{x}{u} \mp i\tau_c  \right) \right] 
		\:\: , 
		\label{ft.8}
		\eneq
		\noindent
		whose Fourier transforms are
		\begin{eqnarray}
			\tilde{\xi}_{\pm , j} ( \omega ) &=& \int \: d t \: e^{ i \omega t } \: \xi_{\pm , j}  ( t ) = \pm \frac{2 \pi e^{ \mp \omega \tau_c}}{ 1 - e^{ \mp \beta_j \omega} }
			\:\: . 
			\label{ft.9}
		\end{eqnarray}
		We compute   the correlation functions of chiral fields including insertions of chiral current operators  using \cite{Campagnano2016}
		\begin{widetext}
			\begin{eqnarray}\label{ft.5}
				&& g_{ \eta ,\{\eta_n\}_n; j}^{\{\alpha_n\}_n, - \nu } \left( t + \frac{x}{u} ; \{t_n\}_n \right) = 
				\left\langle {\cal T}_K \partial_x \varphi_{j,\eta} \left(  t + \frac{x}{u}  \right) 
				\prod_{ l = 1}^N e^{ i \alpha_l \sqrt{4 \pi \nu} \varphi_{j,\eta_l} ( t_l ) } 
				\right\rangle  =  \\
				&& \qquad \qquad \qquad \qquad \qquad\quad  
				= \frac{\sqrt{\nu\pi}}{2 i u \beta_j } \sum_{l=1}^N\alpha_l
				\coth \left[ \frac{\pi}{\beta_j} \left(  t - t_l + \frac{x }{u} -i \tau_c \sigma_{ \eta \eta_l } (t-t_l)  \right) \right]
				g_{\{\eta_n\}_n ; j}^{ \{\alpha_n\}_n, - \nu } \left( t_1-t_2 \right)
				\;\; , \nonumber
			\end{eqnarray}
			\begin{eqnarray}
				&& g_{ \eta  \eta' ,  \{\eta_n\}_n; j}^{\{\alpha\},- \nu } \left( t + \frac{x}{u} ; t' + \frac{x'}{u} ;  
				\{t_n\}_n \right) =  
				\left\langle {\cal T}_K \partial_x \phi_{j,\eta} \left( t + \frac{x}{u} \right) \partial_{x'} \phi_{j,\eta'} \left(  t'+ \frac{x'}{u}\right) 
				\prod_{l = 1}^N e^{ i \alpha_l \sqrt{4 \pi \nu} \varphi_{j,\eta_l} (  t_l ) } \right\rangle  =  \\
				&& \;\; =  - g_{ \{\eta_l\}_l; j}^{-\nu} \left( \left\{ t_i \right\} \right)
				\sum_{ l , l' = 1}^N \frac{ \alpha_l \alpha_{l'} \nu\pi}{4 u^2 \beta^2_j }
				\coth \left[ \frac{\pi}{\beta_j} \left(  t - t_l + \frac{x  }{u}  -i \tau_c \sigma_{ \eta \eta_l } (t-t_l) \right) \right]   
				\coth \left[ \frac{\pi}{\beta_j} \left( t' - t_{l'} + \frac{x' }{u}  -i \tau_c \sigma_{ \eta' \eta_{l'} } (t-t_{l'}) \right) \right]  \nonumber 
				,
				\label{ft.6}
			\end{eqnarray}
		\end{widetext}
		where
		\beq
		\sigma_{ \eta_1 \eta_2 } ( t_1 - t_2 ) = \eta_2\theta\left(t_1 - t_2\right)-\eta_1\theta\left(t_2 - t_1\right)
		\:\: .
		\label{ft.2}
		\eneq
		All the fields in the above contractions are in equilibrium with the $j$-th reservoir at temperature $\beta_j^{-1}$.

		\section{General expression for the leading corrections to the fixed point conductance}
		\label{General:expression}
		
		In this Appendix we provide a general formula for the leading corrections to 
		the FP  conductance tensors  of the junction, given the corresponding leading boundary 
		operator   and the splitting matrix Eq.\eqref{eq:rho}. Consistently with our analysis in the main text, 
		in full generality, we assume that the  leading boundary 
		operator  is realized as a combination of vertex 
		operators depending on pertinent linear combinations of the fields $\varphi_j ( 0 )$, defined in Eq.\eqref{6}. 
		Also, we assume full symmetry between the various channels, which allows us to 
		assume an over-all constant independent of the specific vertex operator. Thus, we set 
		\beq
		H_B = - 2 h \sum_{ k <  l = 1}^N  \cos \left[ \sqrt{4 \pi} \sum_{j = 1}^N  \alpha^{k , l}_j \varphi_j ( 0 ) \right]
		\:\:\:\: , 
		\label{split.5}
		\eneq
		\noindent
		with $\alpha^{k,l}_j$ being a coefficient that explicitly depends on $g$.  For instance, at the DFP,  consistently with
		the boundary interactions of the junction \eqref{exdis.2} and of the TKM \eqref{eq.10} for $N=3$, we set $\alpha^{k,l}_j = g^{-\frac{1}{2}} \: \{ \delta_{k,j}-\delta_{l,j+1} \}$.  
		
		In our picture, the external reservoirs inject into the leads left-handed modes biased at a voltage $V_j$. We account 
		for this by the shift in Eq.\eqref{junct.9}, which provides an explicit dependence on time   in the  leading boundary  
		operator 
		\begin{equation} \label{eq:HB}
			H_B = - 2 h \sum_{ k <  l}^N \cos \Biggl[ \sqrt{4 \pi} \sum_{j = 1}^N \alpha^{k , l}_j\left( \bar{\varphi}_j ( 0 ) + e \sqrt{\frac{g}{4\pi}} V_j t \right) \Biggr].
		\end{equation}
		\noindent
		The electric current operator takes the form
		\begin{eqnarray}\label{jelE}
			j_{{\rm el} , j } ( x , t ) &=&  e u \sqrt{\frac{g}{\pi}} \: \left[ 
			\partial_x \bar{\varphi}_j (t_+ ) -  \sum_{ j' = 1}^N \rho_{j ', j} \partial_x \bar{\varphi}_j (t_- ) \right]
			\nonumber \\
			&& \quad +  \frac{e^2 g}{2 \pi} \: \sum_{j' = 1}^N \: \left\{ \rho_{ j , j'}-\delta_{j,j'} \right\} V_{j'}
			\quad .
		\end{eqnarray} 
		We have already shown that, within linear response theory, the FP  electric conductance is given by Eq.\eqref{junct.12G}, which is 
		perfectly consistent with the result in Eq.\eqref{jelE}. Due to 
		the neutrality constraint on the vertex operators, no corrections arise to  first order in the perturbation Eq.\eqref{eq:HB}.
		Using the Keldysh formalism  up to second order, we obtain the expectation value of the electric and thermal currents in the form 
		\begin{widetext}
			\begin{equation}\label{splitjeexpv}
				I_{{\rm el/th} , j} =    h^2   \: \sum_{ \eta_1 , \eta_2 }  \eta_1 \eta_2 \int d t_1 d t_2
				\sum_{k<l} \prod_{j_1=1}^N\left\langle {\cal T}_K  j_{{\rm el/th} , j} ( x , t ) 
				e^{-i\sqrt{4 \pi} \alpha^{k , l}_{j_1} \varphi_{j_1,\eta_1} ( t_1 )  }
				e^{i\sqrt{4 \pi} \alpha^{k , l}_{j_1} \varphi_{j_1,\eta_2} ( t_2 ) }
				\right\rangle e^{ie\sqrt{g}\alpha_{j_1}^{k , l}V_{j_1}\left(t_1-t_2\right)}+{\rm h.c.}
				\:\: . 
			\end{equation}
		\end{widetext}
		In Eq.\eqref{splitjeexpv} $\eta_{1,2}=\pm1$ label the branch of the Keldysh contour and the current operator lies, by convention, on the upper branch ($\eta = +$). 
		
		We assume an homogeneous temperature throughout the system and turn on a potential bias on lead $j_b\ne j$. As we work close to 
		the charge neutrality point, 
		the Seebeck and Peltier coefficients vanish and only an electric current is generated. Using Eqs. \eqref{jelE} and \eqref{ft.5} and their Fourier transforms Eqs.\eqref{ft.9} and \eqref{eq:ftg}, we obtain
		\begin{eqnarray}
			I_{{\rm el} , j }  &=&  \frac{e h^2 \sqrt{g}}{4 \pi} \sum_{ k < l = 1}^N \sum_{\eta_1 , \eta_2 = \pm 1} 
			\eta_1 \eta_2  \sum_{j' = 1}^N  \rho_{j,j'} \alpha_{j'}^{k , l}   
			\times  \label{split.17}  \\
			&&   \sum_{s =\pm 1} s [ \tilde{\xi}_{\eta_1 ,j' } ( 0 ) - \tilde{\xi}_{\eta_2 , j'} ( 0 ) ]
			\tilde{g}^{- \nu}_{\eta_1 \eta_2 ; j} \left(s_2 e \sqrt{g}
			\alpha_{j_b}^{k , l} V_{j_b } \right)\:.  \nonumber 
		\end{eqnarray}  
		\noindent
		Here $\nu=\sum_{m = 1}^N ( \alpha_{m}^{k , l} )^2$.
		Finally, retaining only contributions that are linear in the applied biases, we obtain the simple expression
		\beq
		I_{{\rm el} , j}  = 2 \pi g e^2 \tilde{h}^2\left( 2 \pi k_B T  \right)  
		\frac{\Gamma^2 \left( {\nu}/{2}\right)}{ \Gamma \left(  \nu \right)}  \label{split.21} 
		\sum_{k < l = 1}^N \sum_{j' = 1}^N \rho_{j,j'}  \alpha^{k,l}_{j'}  \alpha_{j_b}^{k , l}
		\:\: ,
		\nonumber 
		\eneq 
		\noindent
		with the dimensionless coupling \mbox{$
			\tilde{h} \left(D \right) = 
			\frac{ h}{D_0}  \left(  \frac{ D }{D_0} \right)^{- 1 +  \frac{\nu}{2} } 
			$}.
		
		We  now examine the thermal current at charge neutrality Eq. \eqref{junct.8}. We start again from Eq. \eqref{splitjeexpv} 
		and set $V_j=0$ in all leads, but allow for a different temperature  $T_j$  on each lead  $j=1,\ldots,N$. 
		We then make use of Eqs. \eqref{gppgmp} and 
		\eqref{ft.8}, as well as of the Fourier transforms Eq. \eqref{ft.9}, to write the expectation value as
		$   I_{{\rm th} , j}= I_{{\rm th} , j}^{(A)}+ I_{{\rm th} , j}^{(B)}
		$, in which
		\begin{eqnarray}
			&& I_{{\rm th} , j} ^{(A)}= \frac{h^2}{4 \pi}  \sum_{ k < l} \sum_{\eta_1 , \eta_2 = \pm 1}  
			\eta_1 \eta_2 \:\sum_{j_a , j_b } \rho_{j , j_a} \rho_{j , j_b} \alpha_{j_a}^{k,l} \alpha_{j_b}^{k,l} 
			\label{IthA}\\
			&&\quad\times    \tilde{\Gamma}_{\eta_1 , \eta_2}^{ \{ \alpha^{k,l}\} } ( 0 )   \int \frac{d \omega}{2 \pi} 
			\left[ \tilde{\xi}_{\eta_1 , j_a} (  \omega ) \tilde{\xi}_{\eta_1 , j_b} ( -  \omega ) +(\eta_1\to\eta_2)
			\right]\,, \nonumber
		\end{eqnarray}
		\begin{eqnarray}
			&& I_{{\rm th} , j}^{(B)} =\frac{h^2}{4 \pi} \sum_{ k < l} \sum_{\eta_1 , \eta_2 = \pm 1} 
			\eta_1 \eta_2 \sum_{j_a , j_b } \rho_{j , j_a} \rho_{j , j_b} \alpha_{j_a}^{k,l} \alpha_{j_b}^{k,l}
			\label{IthB}\\
			&&\quad\times
			\int  \frac{d \omega}{2 \pi}   \tilde{\Gamma}_{\eta_1 , \eta_2}^{ \{ \alpha^{k,l}\}   }  ( \omega  ) 
			\left[ \tilde{\xi}_{\eta_1 , j_a} (  \omega ) \tilde{\xi}_{\eta_2 , j_b} ( -  \omega ) + (j_a \leftrightarrow j_b)  \right]
			\:\: . \nonumber
		\end{eqnarray}
		The function $\tilde{\Gamma}$ 
		originates the contractions of the derivative of the fields $j_a$ and $j_b$ with the corresponding perturbation at second order. It
		is labeled by the coefficients \mbox{$\{\alpha_{1}^{k,l},\ldots,\alpha_{N}^{k,l}\}$} and stands for
		\beq
		\tilde{\Gamma}_{\eta_1 , \eta_2}^{ \{ \alpha^{ k , l } \}}  ( t ) = \prod_{m=1}^N
		g_{\eta_1 \eta_2;m}^{ - ( \alpha_{m}^{k,l} )^2 }(t)
		\:\: . 
		\label{split.18}
		\eneq
		By going along the derivation of Appendix \ref{vertex_keldysh}, it is easy to verify the following identities
		\begin{equation}
			\sum_{\sigma=\pm}\tilde{\xi}_{\pm , j_a} (\sigma  \omega ) \tilde{\xi}_{\pm , j_b} (-\sigma \omega )=
			-\frac{2\pi^2\cosh\frac{(\beta_a-\beta_b)\omega}{2}}{\sinh\frac{\beta_a\omega}{2}\sinh\frac{\beta_b\omega}{2}},
			\label{xipp}
		\end{equation}
		\begin{eqnarray}
			&&\sum_{\eta=\pm}\tilde{\xi}_{\eta , j_a} (\pm\eta \omega ) \tilde{\xi}_{-\eta , j_b} (\mp\eta\omega )
			\label{xipm}\\
			&&\;=\sum_{\sigma=\pm}\tilde{\xi}_{+ , j_a} (\sigma  \omega ) \tilde{\xi}_{+ , j_b} (-\sigma \omega )
			\nonumber
			\mp 2\pi\left[\tilde{\xi}_{\pm, j_a} ( \omega )+\tilde{\xi}_{\pm, j_b} ( \omega )\right] \; , 
		\end{eqnarray}
		\noindent
		with $\beta_a = (k_B T_a )^{-1}$.  
		Using Eq. \eqref{xipp}, together with the relation
		\beq\label{Gammapm}
		\tilde{\Gamma}_{\eta_1 , \eta_2}^{  \{ \alpha^{k , l}\} }    
		(\omega=0)=  \tilde{\Gamma}_{- \eta_1 , - \eta_2}^{ \{ \alpha^{k , l}\} }  (\omega=0)\;,
		\eneq
		one concludes that Eq. \eqref{IthA} is identically zero. In addition to that, we can exploit Eq. \eqref{xipm} to rewrite Eq.\eqref{IthB} in the simpler form
		\beq
		I_{{\rm th} , j} =  I_{{\rm th} , j}^{ (B)} =    h^2 \sum_{ k < l}\sum_{j_a , j_b }
		\rho_{j , j_a} \rho_{j , j_b} \alpha_{j_a}^{k,l} \alpha_{j_b}^{k,l} \mathcal{A}_{j_a}^{ \{ \alpha^{k , l}\} }   \;,
		\eneq
		with
		\beq\label{Aj}
		\mathcal{A }_{j}^{ \{ \alpha^{k , l}\} } =\int \frac{d\omega}{2 \pi}\left[\tilde{\Gamma}_{-,+}^{ \{ \alpha^{k , l}\}  }(\omega) 
		\tilde{\xi}_{-,j} (\omega)
		-  \tilde{\Gamma}_{+,-}^{ \{ \alpha^{k , l}\}  }(\omega) \tilde{\xi}_{+,j} (\omega)
		\right].
		\eneq
		Using the explicit Fourier transforms Eqs. \eqref{eq:ftg} and \eqref{ft.7}, together with \eqref{ft.9} and the symmetry of
		the system under exchange of wires, it is possible to more explicitly rewrite the integral Eq.\eqref{Aj} as
		\begin{eqnarray}\label{Aexpli}
			&&\mathcal{A}_{j}^{ \{ \alpha^{k , l}\}  }=
			-\int d\omega_1\ldots d\omega_{N}
			\frac{\sinh\frac{\sum_{ s \ne j}(\beta_j-\beta_{s})\omega_l}{2}}{\sinh\frac{\beta_{j}\omega_j}{2}}
			\\
			&&\; \times d_{j}^{-(\alpha_j^{kl})^2}(\omega_j-\omega_1\ldots\hat{\omega}_j\ldots-\omega_{N})
			\prod_{m\ne j}d_{m}^{-(\alpha_m^{kl})^2}(\omega_m) . \nonumber
		\end{eqnarray}
		Here the $\hat{}$ symbol denotes omission. It is now easy to verify that the thermal current vanishes in the 
		absence of temperature gradients, i.e., whenever $\beta_1=\ldots=\beta_N=\beta$.
		
		Focusing on wire $j_0$ (which implies no loss of generality due to the symmetry of the system) and 
		assuming that all the remaining $N-1$ subsystems are in equilibrium at the same temperature $1/\beta$,
		it is then convenient to define the function
		\beq
		\tilde{\Gamma}_{\eta_1,\eta_2}^{ \{ \alpha^{ k , l }; \hat{j}_0\}}   ( t ) = \prod_{m\ne j_0}
		\tilde{g}_{\eta_1 \eta_2;m}^{ - ( \alpha_{m}^{k,l} )^2 }(t)
		\:\: ,
		\label{Gammafew}
		\eneq
		where we have denoted by $\{\alpha;\hat{j}_0\}$ the set $\{ \alpha^{ k , l }\} $, from which the element $ \alpha^{ k , l }_{j_0} $
		has been removed. One can invert the Fourier transform \eqref{eq:ftg} and see directly from the definition \eqref{gppgmp} that
		\beq
		\tilde{\Gamma}_{\mp\pm}^{\{ \alpha^{k , l} ; \hat{j}_0\} }(\omega)=
		e^{ \pm\frac{\beta \omega}{2}} d_j^{-\sum_{m\ne j_0}(\alpha^{k,l}_m)^2}(\omega)
		\eneq
		With the aid of this expression, in the presence of a small temperature bias \mbox{$\beta \to \beta_{j_0}=
			\beta+\delta \beta_{j_0}$} on lead $j_0$ only, the integral $\mathcal{A}_{j}^{\{ \alpha^{k , l}  \}}$ in Eq.\eqref{Aj} is conveniently rewritten as
		\begin{eqnarray}
			&&\mathcal{A}_{j}^{ \{ \alpha^{k , l}\}  }= \sum_{\eta=\pm}\int \frac{d\omega d\omega_0}{2\pi\eta} 
			\tilde{g}^{-\alpha_{j_0}}_{\eta,-\eta}(\omega_{0}-\omega)
			\tilde{\Gamma}_{\eta,-\eta}^{\{ \alpha^{k , l}  ; \hat{j}_0\}}(\omega)
			\tilde{\xi}_{\eta,j}(\omega_{0})\nonumber\\
			&&\; \approx  -  \delta \beta_{j_0} \sum_{\eta=\pm}\int d\omega d\omega_0
			\frac{\omega d_{j_0}(\omega)}{\sinh\frac{\beta\omega_0}{2}}
			d_j^{-\sum_{m\ne j_0}(\alpha^{k,l}_m)^2}(\omega_0-\omega) \nonumber
		\end{eqnarray}
		Gathering all the contributions, the thermal current response in leg $j$ under a change of temperature in leg $j_0$ takes the form
		\begin{equation} \label{split.22}
			I_{{\rm th} , j} = \frac{\pi^2 k_B}{\beta}\sum_{ k < l}\sum_{j_a , j_b }
			\frac{ \rho_{j , j_a} \rho_{j , j_b}  \alpha_{j_a}^{k,l} \alpha_{j_b}^{k,l}\tilde{h}^2\left( \frac{2\pi}{\beta}\right)\delta T_{j_0} 
				\Phi_{j_0}^{\{ \alpha^{k ,l}  \}}}{\Gamma\left[(\alpha^{kl}_{j_0})^2\right]\Gamma\left[\sum_{r\ne j_0}(\alpha^{kl}_{r})^2\right]}
		\end{equation}
		with  $\delta T_{j_0} =  T_{j_0}-T $, and 
		\begin{eqnarray} \label{Phigen}
			&&  \Phi_{j_0}^{ \{ \alpha^{k , l}\}  } =
			\int dz dw 
			\frac{w}{\sinh \left( \pi z \right)} \\
			&&\;\; \times \left| \Gamma \left( \frac{ ( \alpha_{j_0}^{k,l})^2}{2} + i  w \right) 
			\Gamma \left( \frac{\sum_{m \neq j_0} ( \alpha_m^{k,l})^2 }{2} + i (z - w )  \right)\right|^2
			\nonumber
		\end{eqnarray}
		\noindent
		With  Eqs.(\ref{split.21},\ref{split.22}) one can compute  the explicit form of the coefficients 
		entering the leading corrections to the fixed point value of the CCT and of the HCT.
		In order to improve the perturbative result, the running coupling strengths may be inserted instead of 
		the bare ones in front of the right-hand side of   Eqs.(\ref{split.21},\ref{split.22}).
		Within a straightforward generalization of the formalism we develop here, it is possible to address the case 
		in which the  leading boundary 
		operator is made out of a linear combination of vertex operators with different coefficients. 
		Using the values of $ \alpha_m^{k,l}$ from Section \ref{discfip_3} we obtain the expressions \eqref{Gpert},
		\eqref{Kpert} and \eqref{n3.2}.
		
		Finally, we quote the identities
		\begin{eqnarray}\label{dfp.2.13}
			&&
			\Gamma\left(\frac{1}{2}+iz\right)\Gamma\left(\frac{1}{2}-iz\right)=
			\frac{\pi}{\cosh\left(\pi z\right)} \;,     \\
			&& \int dz \frac{\pi}{\sinh ( \pi z )\cosh ( \pi (z-w) )}=\frac{2\pi w}{\cosh(\pi w)}  \, , \nonumber\\
			&& \Gamma\left(z^*\right)=\Gamma\left(z\right)^* \;,\quad 
			\int \: d w \: \frac{w^2}{\cosh^2 ( \pi w )} =\frac{1}{6 \pi}  \nonumber
			\:\: 
		\end{eqnarray}
		which are used to obtain an explicit expression of the Lorenz ratio in the main text. 
		
		\section{Boundary renormalization of the Topological Kondo model}
		\label{app:RGTKM}
		
		In this Appendix  we provide more details about the renormalization of the boundary term Eq.\eqref{eq.10}. 
		To begin with, we recall that the differences between the running couplings 
		associated to the   $J_{k , l}$ are systematically washed out along the RG trajectories \cite{Beri2012}. This is encoded 
		in the perturbative RG equations 
		for the running couplings   ${\cal J}_{k,l} (D) \equiv J_{k,l} ( D )^{- 1 +\frac{1}{g}}$.
		Within $\epsilon $ expansion, with $0 < \epsilon (=1-g^{-1} ) \ll 1$, these are given by   \cite{Cardy1996,giuliano2018,giuliano2019,Kane2020}
		\beq
		\frac{d {\cal J}_{k,l}}{d \ln \left( \frac{D_0}{D} \right)} 
		= \epsilon {\cal J}_{k,l} + 2 \sum_{ a ( \neq k , l) = 1}^N {\cal J}_{k,a} {\cal J}_{a,l} 
		\:\:\:\: ,
		\label{ren.3}
		\eneq
		\noindent
		Based on the above observation, 
		in order to discuss the flow toward the $D^{N-1}$ fixed point, we
		consider the symmetric version of Eq.\eqref{ren.3}, in which all 
		the ${\cal J}_{j,k}$ are taken equal to each other and equal to ${\cal J}$, so to recover 
		a single RG equation
		\beq
		\frac{d {\cal J}}{d \ln \left( \frac{D_0}{D} \right)} = 
		\epsilon {\cal J} + 2 (N-2) {\cal J}^2 
		\:\:\:\: , 
		\label{ren.4}
		\eneq
		which is solved by 
		\beq
		{\cal J} ( D ) = \frac{\epsilon {\cal J}_0  \left( \frac{D_0}{D} \right)^\epsilon }{\epsilon + 2 (N-2) {\cal J}_0  - 2 (N-2 ) {\cal J}_0
			\left( \frac{D_0}{D} \right)^\epsilon}
		\;\;\;\; , 
		\label{ren.5}
		\eneq
		\noindent
		with ${\cal J}_0 = {\cal J} (D=D_0)$. Identifying  the running energy scale $D$ in Eq.\eqref{ren.5}
		with (the Boltzmann constant times) the temperature $T$ we see that, due to the relevance of 
		the boundary interaction for $g \geq 1$, the system crosses over to the strongly 
		coupled regime at a crossover temperature  $T_*$, determined by the condition that 
		the denominator of the right-hand side of Eq.\eqref{ren.5} becomes equal to 0 at 
		$D = D_* = k_B T_*$. Accordingly, we obtain 
		\beq
		\epsilon^{-1} \left\{ \left( \frac{D_0}{k_B T_*} \right)^\epsilon  - 1 \right\} = 
		\frac{1}{2 (N-2) {\cal J}_0 }
		\:\:\:\: , 
		\label{ren.6}
		\eneq
		\noindent
		which, for $\epsilon \to 0$, reduces to the ``standard'' Kondo result 
		\beq
		\ln \left( \frac{D_0}{k_B T_*} \right) = 1 + \frac{1}{2 (N-2) {\cal J}_0 }
		\:\:\:\: . 
		\label{ren.7}
		\eneq
		\noindent
		At scales $T \leq T_*$ the system enters the strongly coupled regime. 
		
		In the presence of a nonvanishing phase $\chi$  Eq.\eqref{eq.10} generalizes to the 
		set of RG equations for the running parameters ${\cal J} ( D ) , \chi ( D )$ given by  \cite{Giuliano2009}
		
		\begin{eqnarray}
			\frac{ d {\cal J} }{d \ln \left( \frac{D_0}{D} \right)} &=& \epsilon {\cal J} + 2 \cos ( \chi ) {\cal J}^2 \nonumber \\
			\frac{d \chi}{d \ln \left( \frac{D_0}{D} \right)} &=& - 2  \sin ( \chi ) {\cal J}
			\:\:\:\: . 
			\label{rengcom.2}
		\end{eqnarray}
		\noindent
		Eqs.\eqref{rengcom.2} typically emerge when considering the RG approach to the boundary interaction at a junction of three bosonic, one-dimensional 
		interacting system. To construct the splitting matrix ${\bf \rho}$, we first
		minimize the TKM boundary interaction Hamiltonian in Eq.\eqref{eq.10}, 
		rewritten in terms of  $\Phi $ and of the $ \xi_a$ defined in Eq.\eqref{ren.8}.
		Then, we  impose Dirichlet boundary conditions by pinning $\xi_a ( 0 )$, $\forall a = 1 , \ldots , N-1$ (note that $\Phi$
		decouples from the boundary interaction, due to the total charge conservation \cite{Oshikawa2006}).

		As discussed in sec. \ref{phadiag}, the system possesses two strong-coupling FPs in certain parameter ranges, when the boundary potential created by 
		the interaction terms dominates over the kinetic part of the Hamiltonian. The corresponding conformal boundary conditions  are, 
		therefore, obtained by imposing Dirichlet boundary
		conditions at $x=0$ on the fields $\xi_1 ( x ) , \xi_2 ( x )$ in Eq.\eqref{eq.10},
		which take values lying at the sites of a triangular/hexagonal lattice. The operator driving the system away from a fixed point described by Dirichlet 
		boundary conditions on the $\xi_a$ corresponds to a combination of instanton operators encoding  ``jumps'' between sites of the lattice of the minima.
		
		The $D^2$ FP, or topological Kondo FP, extensively discussed in literature, is described by a triangular lattice. The $\hat{D}^2$ FP, instead, 
		is described by requiring that the values of $( \xi_1 ( 0 ) , \xi_2 ( 0 ) )$ minimizing the boundary potential span a honeycomb lattice 
		(determined by two interpenetrating ``rotated'' triangular lattices \cite{Yi1998,Giuliano2009}). Due to the reduced distance between the  sites 
		of the lattice of the minima, "short instanton" operators emerge, with a lower dimension than the boundary operators at the $D^2$FP. 
		The boundary conditions at the $\hat{D}^2$FP are the same as at the $D^2$FP, so, we readily conclude that ${\bf \rho}_{\hat{D}^2} = {\bf \rho}_{D^2}$.
		The leading boundary interaction at the $\hat{D}^2$FP is given by \cite{Yi1998,Giuliano2009}
		\beq
		\hat{\tilde{H}}_{{\rm TK} , 2}  = - 2 \hat{h}  \sum_{ j = 1}^3 \: e^{ i \frac{4 \sqrt{ \pi g} }{3 \sqrt{3}} ( 2 \tilde{\varphi}_j ( 0 ) - 
			\tilde{\varphi}_{j+1} ( 0 ) - \tilde{\varphi}_{j-1} ( 0 ) ) } \tau^+ 
		+ {\rm h.c.} 
		\:\:\:\: .
		\label{rengcom.6}
		\eneq
		Following \cite{Yi1998,Giuliano2009}, in Eq.\eqref{rengcom.6} we have 
		introduced an effective isospin operator $\vec{\tau}$. We have done so in order to 
		take into account that the honeycomb lattice of minima at the $\hat{D}^2$FP is 
		made out of the interpenetration of two inequivalent triangular lattices, say 
		${\bf A}$ and ${\bf B}$. Any short instanton originating from sublattice ${\bf  A}$ 
		ends into one of the three nearest neighboring sites of sublattice ${\bf B}$, and vice
		versa. Thus, having arbitrarily associated the eigenvalue $+1$ of $\tau_z$ to states 
		living over sublattice ${\bf A}$ and the eigenvalue $-1$ to states living over 
		sublattice ${\bf B}$, we recover the final expression in Eq.\eqref{rengcom.6}. 
		
		\bibliography{long_paper_revision.bib}
		
	\end{document}